\documentclass[sigconf,nonacm]{acmart} 

\usepackage{listings}
\usepackage[table]{xcolor} 
\usepackage{xspace}

\usepackage{enumitem}
\setlist{leftmargin=!, labelwidth=1.2em} 

\usepackage{multirow}
\usepackage{calc}
\usepackage{hyperref}

\usepackage[utf8]{inputenc} 

\usepackage{tcolorbox}
\usepackage{stfloats}
\usepackage{wrapfig}
\usepackage{graphicx}

\usepackage{tabularx}
\usepackage{relsize} 
\usepackage{trimclip} 
\usepackage{natbib} 

\usepackage{etoolbox} 
\usepackage{makecell} 
\usepackage{anyfontsize} 

\usepackage{ragged2e}

\newcommand{\todo}[1]{\textbf{\color{red}{#1}}}

\newcommand{\eg}{\textit{e.g.\@}\xspace}

\newcommand{\ie}{\textit{i.e.\@}\xspace}

\newcommand{\notarxiv}[1]{}

\newcommand{\titletext}{Professional Software Developers Don't Vibe, They Control:\\AI Agent Use for Coding in 2025} 

\newcommand{\myshorttitle}{Professional Software Developers Don't Vibe, They Control: AI Agent Use for Coding in 2025}

\graphicspath{{images/}}

\makeatletter
\renewenvironment{quotation}
  {\list{}{\leftmargin=1em \rightmargin=1em \topsep=0pt \parsep=0pt}%
   \item\relax}
  {\endlist}
\makeatother

\newcommand{\topquote}[2]{
\begin{quotation}
\emph{#1} \hfill\mbox{- #2}
\end{quotation}
\vspace{0.5em}
}

\definecolor{barcolor}{rgb}{0.247,0.408,0.671}
\definecolor{plotnumcolor}{rgb}{0.624,0.704,0.836}

\newcommand{\barRow}[2]{%
  #1\,&%
  \raisebox{-0.1em}{\tikz \fill[barcolor] (0,0) rectangle (\barscale*#2pt, 0.75em);}%
  \ \textcolor{plotnumcolor}{#2}%
  \\%
}

\newenvironment{hbars}[2]{%
  \footnotesize\centering
    \def\barscale{#2}%
    \renewcommand{\arraystretch}{0.9}%
    \setlength{\tabcolsep}{0pt}%
    \begin{tabular}[t]{r|l}
      \hiderowcolors
      \multicolumn{2}{c}{\textbf{#1}}\\[0.3em]
}{%
    \end{tabular}%
}

\newlength{\origcolumnsep}
\newlength{\origintextsep}

\definecolor{CodeGrayColor}{RGB}{83, 83, 89}

\newcommand{\scode}[1]{{\small\texttt{#1}}}
\newcommand{\myparaskip}{\smallskip}
\newcommand{\myparaheadstyle}[1]{\textbf{\textit{#1.}}}
\newcommand{\mypara}[1]{{\myparaskip\noindent\myparaheadstyle{#1}\xspace}}

\newcommand{\mytakeaway}[1]{\begin{tcolorbox}[
  width=\linewidth,
  colframe=black!100,arc=1.5pt,boxrule=0.3pt,
  colback=gray!10,
  left=0.5em,right=0.5em,top=0.5em,bottom=0.5em, 
  before skip=0.5em,after skip=0.5em          
]#1\end{tcolorbox}}

\newcommand{\promptbox}[1]{\begin{tcolorbox}[
  width=\linewidth,
  colframe=gray!33,arc=1.5pt,boxrule=0.3pt,
  colback=gray!2,
  left=0.25em,right=0.25em,top=0.1em,bottom=0.1em, 
  before skip=0em,after skip=0em          
]#1\end{tcolorbox}}

\definecolor{basicColor}{rgb}{0,0,0}
\definecolor{backgroundColor}{rgb}{.97,.97,.97}
\definecolor{identifierColor}{rgb}{0.0,0.0,0.0}
\definecolor{keywordColor}{rgb}{.22,.49,.13}
\definecolor{stringColor}{rgb}{.67,.19,.16}
\definecolor{numberColor}{rgb}{.23,.53,.14}
\definecolor{operatorColor}{rgb}{.61,.18,.96}
\definecolor{commentColor}{rgb}{0.5,0.5,0.5}
\lstdefinestyle{notebookish}{
  frame=single,
  backgroundcolor=\color{backgroundColor}, 
  columns=[l]fullflexible, 
  basicstyle=\linespread{1.0}\footnotesize\ttfamily\color{basicColor}, 
  breaklines=true, 
  keepspaces=true,
  breakatwhitespace=true,
  xleftmargin=1\parindent,
  xrightmargin=1\parindent,
  language=Python,
  showstringspaces=false,
  morekeywords={[1]True,False,None},
  otherkeywords={0,1,2,3,4,5,6,7,8,9,=,+,-,*,/}, 
  morekeywords={[3]0,1,2,3,4,5,6,7,8,9}, 
  morekeywords={[4]=,+,-,*,/}, 
  keywordstyle=\bfseries\color{keywordColor},
  keywordstyle=[2]\color{keywordColor}, 
  keywordstyle=[3]\color{numberColor}, 
  keywordstyle=[4]\bfseries\color{operatorColor},
  commentstyle=\color{commentColor},
  identifierstyle=\color{identifierColor},
  stringstyle=\color{stringColor},
  extendedchars=true,
  literate={∞}{{$\infty$}}1,
  escapeinside={<@}{@>}, 
}
\definecolor{mycolor}{rgb}{0.1,0.4,0.1}
\definecolor{lightblue}{rgb}{0.9,0.9,1.0}

\newcommand{\studyOne}[0]{Observations\xspace}
\newcommand{\studyTwo}[0]{Survey\xspace}
\newcommand{\numStudyOne}[1]{#1x \studyOne}
\newcommand{\numStudyTwo}[1]{#1x \studyTwo}

\newcommand{\packageURL}[0]{\url{https://tinyurl.com/devs-using-agents-2025}\xspace}

\newcommand{\newS}[1]{%
  \ifstrequal{#1}{99}{S1}{%
  \ifstrequal{#1}{100}{S39}{%
  \ifstrequal{#1}{101}{S86}{%
  \ifstrequal{#1}{103}{S97}{%
  \ifstrequal{#1}{104}{S102}{\todo{NOT A REMAPPED SURVEY RESPONDENT}}%
  }}}}%
}

\begin{document}

\title[\myshorttitle]{\titletext}


\author{Ruanqianqian (Lisa) Huang}
\authornote{Denotes equal contribution.}
\authornote{Work done at UC San Diego.}
\email{lisahuang@cornell.edu}
\affiliation{\institution{Cornell University} \country{USA}}

\author{Avery Reyna}
\authornotemark[1]
\email{avery.reyna16@gmail.com}
\affiliation{\institution{Independent} \country{USA}}

\author{Sorin Lerner}
\email{sorin.lerner@cornell.edu}
\affiliation{\institution{Cornell University} \country{USA}}

\author{Haijun Xia}
\email{haijunxia@ucsd.edu}
\affiliation{\institution{UC San Diego} \country{USA}}

\author{Brian Hempel}
\email{bhempel@ucsd.edu}
\affiliation{\institution{UC San Diego} \country{USA}}

\renewcommand{\shortauthors}{Huang and Reyna et al.}

\begin{abstract}
The rise of AI \emph{agents} is transforming how software can be built. 
The promise of agents is that developers might write code quicker, delegate multiple tasks to different agents, and even write a full piece of software purely out of natural language. In reality, what roles agents play in professional software development remains in question.
This paper investigates how \emph{experienced} developers use agents in building software, including their motivations, strategies, task suitability, and sentiments.
Through field observations (N=13) and qualitative surveys (N=99), we find that while experienced developers value agents as a productivity boost, they retain their agency in software design and implementation out of insistence on fundamental software quality attributes, employing strategies for controlling agent behavior leveraging their expertise.
In addition, experienced developers enjoy working with agents as source of collaboration rather than complete delegation given their judgment for task suitability.
Our results shed light on the value of software development best practices in effective use of agents, suggest the kinds of tasks for which agents may be suitable, and point towards future opportunities for better agentic interfaces and agentic use guidelines.

\end{abstract}

\notarxiv{



\received{20 February 2007}
\received[revised]{12 March 2009}
\received[accepted]{5 June 2009}
}

\maketitle

\section{Introduction}
\label{sec:intro}

\topquote{I've been a software developer and data analyst for 20 years and there is no way I'll EVER go back to coding by hand. That ship has sailed and good riddance to it.}{Developer in Our Survey (S28)}

AI is rapidly changing the practice of programming.
Already, about half of professional software developers are using AI tools daily~\cite{stackoverflow2025survey}.
Although originally introduced into coding in 2021 as a super-charged autocomplete~\cite{GitHubCopilot}, recent large language models (LLMs) can now work \emph{agenticly}, accessing, modifying, and testing whole codebases in autonomous, step-by-step actions.
Some online users claim to use dozens of agents at once to autonomously construct massive software (\eg,~\cite{thermo2025dozenAgents, yegge2025dozenAgents}), a claim so intriguing but potentially incredulous that it is parodied~\cite{ivibemorethanyoudotcom}.
But in contrast, anecdotally, we sometimes hear from people that they tried coding with agents and it didn't work out so well.
\emph{What is really happening?}

Human studies of agentic coding are emerging but sparse. 
A randomized trial found that experienced open source maintainers were slowed down by 19\% when allowed to use AI~\cite{becker2025early2025}, and an agentic system deployed in an issue tracker saw only 8\% of its invocations resulting in complete success~\cite{takerngsaksiri2025hula}.
While this suggests agents are less useful than expected, a quarter of professional developers nevertheless report that they already use AI agents at least weekly~\cite{stackoverflow2025survey}.
A few investigations~\cite{sarkar2025vibe,fawzy2025vibeCoding,pimenova2025goodVibrations,geng2025studentVibe} studied ``vibe coding'' or \emph{vibing}~\cite{karpathy2025vibecoding}, a particular form of agent use that emphasizes ``flow and joy'' by trusting the AI instead of carefully reviewing the generated code~\cite{pimenova2025goodVibrations, karpathy2025vibecoding}. 
But vibing may produce lower quality code~\cite{fawzy2025vibeCoding}
and may not be how experienced developers use agents.
\emph{How, then, do experienced developers create software with AI agents?}

This paper is an attempt to gain insight into the current practice of agentic coding by experts.
Compared to prior work, we (a) do not limit the investigation to just vibe coding, and (b) examine \emph{experienced} developers only, in the hopes that they can be insightfully critical of the agentic tools in real-world use.
We present a two-part study---13 field observations and a survey of 99 experienced developers---hoping to answer four research questions (RQs):

\newcommand{\RQ}[1]{\textbf{RQ{#1}}}

%
%
%

\begin{itemize}[labelwidth=7.9em,labelsep=0.35em,align=left]
    \item [\RQ{1 Values.}] What do experienced developers care about when integrating agents into their workflow?
    \item [\RQ{2 Strategies.}] What strategies do experienced developers employ when developing software with agents?
    \item [\RQ{3 Suitability.}] What are software development agents suitable for, and when do they fail?
    \item [\RQ{4 Sentiments.}] What sentiments do experienced developers feel when using agentic tools?
\end{itemize}

Our most salient finding is that, indeed, \textbf{professional developers do not vibe code.
Instead, they carefully \emph{control} the agents through planning and supervision.}
Specifically, 
they are looking for a productivity boost while still valuing software quality attributes (RQ1), they plan before implementing and validate all agentic outputs (RQ2), they find agents suitable for well-described, straightforward tasks but not complex tasks (RQ3), and yet they generally enjoy using agents as long as they are in control (RQ4).
We discuss the implications of these results, and call for future work that inspects the above practices in depth and evaluate their impact on the successful use of agents.



\section{Methods}
\label{sec:methods}

We define \emph{experienced developers} as those with at least three years of professional development experience,
and \emph{agentic tools} or \emph{agents} as AI tools integrated into an IDE or a terminal that can manipulate the code directly (\ie, excluding web-based chat interfaces).

We study the RQs through a two-part study: an observational study (\autoref{subsec:methods-study-one}) and a survey (\autoref{subsec:methods-study-two}).
The field observations focused on experienced developers who adopted AI tools due to workplace requirements, while the survey targeted experienced developers who demonstrated interest in AI-assisted software development on GitHub. 
Upon approval from our institutional review board, we ran both parts between August and October 2025.
We share the study materials (artifact) at \packageURL.

\subsection{Part 1: Field Observations}\label{subsec:methods-study-one}

\mypara{Participants}
We recruited 13 participants (1 female, 12 male) via networks (12) and snowball sampling (1), all of whom had $\ge$3 years of verifiable professional software engineering experience ($range=[3, 25], median=9, sd=6.5$), prior agentic tool experience, and ability to demonstrate a realistic task to be done with agents.
\autoref{tab:observation-participants} details their agentic tool use, model use, and tasks.

\mypara{Procedure}
Each session included a 45-minute observation and a 30-minute semi-structured interview, recorded on Zoom and run by the first two authors.
Participants worked on self-selected tasks (more below) using their preferred setup while thinking aloud.
We asked questions about workflows and tasks throughout to gain additional insights, following prior practices~\cite{groundedJupyter}.
The interview dived into our research questions and necessary observation follow-ups.
Upon completion, each participant received a \$100 USD gift card.
The supplement includes procedural details.

\mypara{Tasks}
Participants brought ongoing or from-scratch tasks for work or side projects using their usual agentic workflows; tasks need not be completed during the study.
For work projects, five worked on production software tasks (P2, P3, P6, P7, P13) and three demonstrated exploratory R\&D tasks (P4, P5, P10); the remaining five demonstrated non-work projects that were open-source collaborations (P1, P9, P12) or personal side projects (P8, P12).
Five participants (P1, P4, P5, P9, P12) reported their tasks were outside their professional domains (the ``Expert?'' column in \autoref{tab:observation-participants}).

\begin{table}[b]
\newcommand{\claudecode}{CC\xspace}
\newcommand{\windsurf}{W\xspace}
\newcommand{\kilocode}{KC\xspace}
\newcommand{\terragon}{T\xspace}
\newcommand{\sonnet}{S\xspace}
\newcommand{\gemini}{Gem\xspace}
\newcommand{\gpt}{G\xspace}
\centering
\footnotesize
\setlength{\tabcolsep}{1.3pt}
\caption{Observational study participants, agentic tools \& models used during the study, and tasks. ``YoE'' = years of professional software development experience; ``Expert?'' = whether a task is within one's expertise.
}
\label{tab:observation-participants}
\rowcolors{1}{}{gray!10}
\begin{tabularx}{\linewidth}{cc>{\raggedright\arraybackslash}p{1.17cm}>{\raggedright\arraybackslash}p{1.1cm}cX}

\hiderowcolors
\toprule
\textbf{ID} & \textbf{YoE} & \textbf{Tool(s)} & \textbf{Model(s)} & \textbf{Expert?} & \textbf{Task} \global\rownum=0\relax \\
\showrowcolors
\midrule
P1 & 5 & \claudecode{}, \windsurf{} & \sonnet{}4, o3, \gemini{}2.5P & & Building a login system for data labeling platform \\
P2 & 9 & \windsurf{} & \sonnet{}3.7, \gpt{}5 & \checkmark & Planning to implement a React app  \\
P3 & 10 & Cursor & auto & \checkmark & Improving a prototype for AI safety \\
P4 & 6 & Cursor & \sonnet{}3.5, \gpt{}5 & & Improving a user tutorial for a dashboard\\
P5 & 9 & Cursor & \gpt{}5, \sonnet{}4 & & Debugging a UI for object detection \\
P6 & 11 & Cursor & \sonnet{}4, \gpt{}5 & \checkmark & Generating an API and relevant tests \\
P7 & 3 & GH Copilot & \sonnet{}4 & \checkmark & Building an ML detection pipeline \\
P8 & 6 & GH Copilot & \sonnet{}3.5, \gpt{}4o & \checkmark & Building app that visualizes file histories \\
P9 & 3 & \kilocode{}, \terragon{} & \sonnet{}4 & \checkmark & Migrating Radix UI assets to Base UI \\
P10 & 15 & \claudecode{}, Cursor & \sonnet{}4, \gpt{}5 & \checkmark & Debugging data pipeline \& improving UI \\
P11 & 25 & \claudecode{}, Codex & \sonnet{}4, codex1 & \checkmark & Building a Ruby-based card game \\
P12 & 9 & Cursor & \sonnet{}4 & & Improving a chatbot \& refactoring code \\
P13 & 20 & \claudecode{} & \sonnet{}4.5 & \checkmark & Debugging a production testing suite \\
\bottomrule
\hiderowcolors
\multicolumn{2}{r}{\textit{Tools:}}&\multicolumn{4}{l}{\textit{\claudecode = Claude Code, \windsurf = Windsurf, \kilocode = Kilo Code, \terragon = Terragon}}\\
\multicolumn{2}{r}{\textit{Models:}}&\multicolumn{4}{l}{\textit{\sonnet = Sonnet, \gemini = Gemini, \gpt = GPT.}}\\
\end{tabularx}
\end{table}

\subsection{Part 2: Invited Qualitative Survey}\label{subsec:methods-study-two}

\mypara{Survey Structure}
We designed a 15-minute qualitative survey with all questions required (except for the last on optional comments), iteratively refined via three institutional pilots (14 participants, excluded from final analysis), where completing participants entered a drawing for four \$100 USD gift cards.
The survey contains three sections 
(exact wording of first two in supplement; full survey in artifact).
Section I asked 11 questions grounding participants' last agentic AI use, probing their (1) task, (2) general vs. (3) agentic priorities, (4) tool choices and (5) rationales, (6) prompting strategies, (7) multi-agent usage, (8) code modification frequency, (9) task success and (10) agent suitability, and (11) enjoyment.
Section II posed two open-ended questions regarding tasks they prefer to perform (12) without and (13) with agentic assistance based on their overall software development experience in the past year.
Section III collected demographics and software development experience.





\mypara{Survey Participants}
Following prior approach~\cite{liang2024survey}, we recruited GitHub users from the following AI-relevant repositories to reach potential enthusiasts
(full list in the supplement):
two popular tools (Cursor, Claude Code), three tools ranked among top 20 on OpenRouter~\cite{OpenRouterRankings} (Kilo Code, Cline, RooCode), AI/ML and development frameworks, open-source language models, and ones tagged \scode{agentic} or \scode{agentic-coding} with $\ge$500 stars.
Using GitHub's GraphQL API, we retrieved users by querying  commits, pull requests (PRs), issues, stargazers, and forkers in these repositories within the past year.
The deduplicated set union yielded 4,141 users, surveyed via unique email links via Qualtrics, resulting in 249 responses (6\% rate, comparable to prior work~\cite{liang2024survey, liang2023developerdesigndecisions}).
We retained 104 qualified users ($\ge$18 years of age, $\ge$3 years of professional experience, and experience with $\ge$1 agentic tool), and after discarding five suspicious AI-generated responses, analyzed the remaining 99.
These respondents (97 male, 1 female, 1 undisclosed) averaged 12.8 years of professional experience ($range=[3, 41], median=10, sd=9.7$), geographically spread across Asia (29), North America (27), Europe (24), South America (13), Africa (5), and Oceania (1).
We translated five non-English responses (via Gemini 2.5 Pro and Claude Sonnet 4) and open-coded the reported tasks and experience domains, finding respondents were majorly full-stack developers (\autoref{fig:survey_tasks_experience_tools_desired_attributes}a), with 96 reporting specific tasks like app building (\autoref{fig:survey_tasks_experience_tools_desired_attributes}b, omitting three who reported general use cases).

\subsection{Data Analysis}\label{subsec:methods-data-analysis}

We performed two stages of data analysis to answer the four RQs.
First, we followed a standard thematic analysis process~\cite{braunUsingThematicAnalysis2006} to analyze the qualitative data and develop initial answers; because of the two-part nature of our study, we conducted open coding separately on each part and developed themes upon all codes from both.
Then, we analyzed specific parts of the data in depth, as some of the emergent themes warranted further investigation.

\mypara{\studyOne}
After all recordings were auto-transcribed by Zoom,
the first two authors conducted open coding of the first two sessions, using field notes as references, to develop the initial code book.
Then, they each coded one half of the remaining sessions, discussing new codes to achieve agreements every two sessions.

\mypara{\studyTwo}
In parallel to analyzing \studyOne, the same two authors conducted open coding of the open-ended survey responses, each analyzing half of the first 30 responses to develop an independent code book.
They then each coded half of the remaining responses, meeting afterwards to discuss notable findings and share sample data of new codes to establish agreements.

\mypara{Initial Joint Thematic Analysis}
Given the large quantity of our qualitative data, rather than computing inter-rater reliability, both coders iteratively agreed on each code creation and instance following best practices~\cite{mcdonald2019reliability}.
Afterwards, the first and last authors used affinity mapping
to cluster semantically similar codes from both study parts, merging clusters as needed to develop a description for each that became a \emph{theme}.
Two emerging themes required further investigation to fully answer our research questions: (for RQ2) developers \emph{control} rather than vibe when using coding agents---but how did they control?
And (for RQ3) developers mentioned using agents for several tasks---but which specific tasks are agents suitable for?
We answered these questions via targeted secondary analyses of the prompts from \studyOne and tasks mentioned in \studyTwo.


\begin{figure*}[b]
\begin{hbars}{(a) Most Recent Task for Survey}{0.7}
  \barRow{Building apps}{53}
  \barRow{Backend}{12}
  \barRow{Automation}{10}
  \barRow{Cloud Computing}{5}
  \barRow{Frontend}{4}
  \barRow{General Exploration}{4}
  \barRow{Refactoring}{4}
  \barRow{System Programming}{4}
  \barRow{Building ML Models}{3}
  \barRow{Generating Agentic Workflow}{3}
  \barRow{Visualization}{3}
  \barRow{\emph{Other}}{3}
\end{hbars}	%
\hfill%
\begin{hbars}{(b) Developer Experience Domain}{0.7}
  \barRow{Full-Stack}{51}
  \barRow{Backend}{21}
  \barRow{AI/ML}{15}
  \barRow{Frontend}{10}
  \barRow{Data Analytics}{5}
  \barRow{Ecommerce}{4}
  \barRow{3D Modeling}{1}
  \barRow{Architecture}{1}
  \barRow{B2B}{1}
  \barRow{Mobile Development}{1}
\end{hbars}
\hfill%
\begin{hbars}{(c) Tools Used}{1}
    \barRow{Claude Code}{58}
    \barRow{GitHub Copilot}{53}
    \barRow{Cursor}{51}
    \barRow{Windsurf}{15}
    \barRow{Codex}{15}
    \barRow{Roo Code}{12}
    \barRow{Gemini CLI (user)}{12}
    \barRow{Cline}{10}
    \barRow{Kilo Code (user)}{9}
    \barRow{Augment Code (user)}{3}
    \barRow{Aider (user)}{3}
    \barRow{\emph{Other (user)}}{22}
\end{hbars}
\hfill%
\begin{hbars}{(d) Desired Quality Attributes w/AI}{3}
    \barRow{Correctness}{20}
    \barRow{Readability}{18}
    \barRow{Modularity}{7}
    \barRow{Performance}{7}
    \barRow{Deployability}{6}
    \barRow{Reliability}{6}
    \barRow{Maintainability}{5}
    \barRow{Scalability}{5}
    \barRow{Relevance}{4}
    \barRow{Usability}{4}
    \barRow{Security}{2}
    \barRow{\emph{Other}}{2}
\end{hbars}
  \caption{In the surveys: (a) most recent agentic task; (b) experience domain; (c) agentic tools used (may report multiple, ``(user)'' = not a provided option and reported via a text box); and (d) software quality attributes desired when working with agents. Numbers may sum to more than 99 because of multiple response or multiple categorization of a single response.
  }
  \label{fig:survey_tasks_experience_tools_desired_attributes}
\end{figure*}

\mypara{Targeted Secondary Analyses}
For our \textbf{prompt analysis}, we open-coded the transcribed \studyOne prompts (\ie, user queries) and plans (\eg, Markdown files) to identify referenced context varieties.
Two researchers co-developed an initial code book on the first session, then each coded half of the remaining sessions, meeting iteratively to refine codes.
We report the top ten of 43 identified context types (\autoref{tab:prompt_context}).
Separately, one researcher recorded word counts and estimated requested steps per prompt/plan by manually counting discrete instructions (\eg, todo list items, imperative verbs, questions), as well as actual executed steps---which could differ from requests due to user interruptions or prior plans---reported in \autoref{tab:prompt_size_and_verification}.
Our \textbf{task suitability analysis} was rather \emph{reflexive}~\cite{braunUsingThematicAnalysis2006}: one researcher first extracted 189 tasks from raw survey responses and merged them into 116 initial codes with another researcher, 
then manually coded the agent's suitability per task code, 
and finally, with another researcher, refined the suitability codes using relevant quotes pulled by OpenAI GPT-5-mini.
The final 89 task codes averaged 8.5 per survey; \autoref{tab:suitability} reports the 59 appearing in at least 5 surveys.
The suitability thresholds were decided as part of the reflexive process (more in \autoref{subsec:results-suitability}).

\subsection{Limitations}

We acknowledge several threats to the validity of our findings.
First, the survey involved a deliberate selection bias towards people positive towards AI, which was desirable for answering RQ2 and RQ3 but perhaps less appropriate for RQ4.
Our 13 field observations also represent a relatively small sample due to the difficulty in recruiting experienced developers who could share work-related tasks when there were no personal projects.
We addressed this threat by augmenting the observations with the large-scale survey of 99 developers.
Second, despite recruitment efforts, 12/13 (92\%) observation participants and 97/99 (98\%) survey respondents self-identified as male, slightly higher than the global stats as of 2024 (91\%)~\cite{JetBrainsDevEcosystem}, limiting gender diversity in our sample. 
Third, our survey recruitment involved scraping public emails from GitHub (similar to a prior survey~\cite{liang2024survey}), introducing a potential self-selection bias toward developers active on public and AI/ML repositories; despite this, respondents reported a wide variety of tasks (\autoref{fig:survey_tasks_experience_tools_desired_attributes}a).
Fourth, it is possible that participants may not have been as expert as they represented; to mitigate, we verified observation participants' web presence for 3+ years of experience (12x by LinkedIn work history, 1x by company website), recruited survey respondents strictly via direct email, and did not disclose any precise experience cutoff in the survey.
Finally, each observation was limited to a single 45-minute session, \ie, we were unable to examine longitudinal use or software development implemented in its full cycle; as a mitigation, we asked both observation and survey participants about their general experience outside of the observed/surveyed tasks.

\section{Results}
\label{sec:results}

Below we report results for each RQ using both parts of the study, referencing \studyOne participants with prefixes of P and \studyTwo ones with~S.



\subsection{RQ1: Values when Using Agents}\label{subsec:results-values}

\topquote{I'm on disability, but agents let me code again and be more productive than ever (in a 25+ year career).}{S22}

We saw a variety of agent use: \studyOne participants used Cursor (6/13) and Claude Code (4/13) the most (\autoref{tab:observation-participants}), whereas \studyTwo participants reported Claude Code (58/99), GitHub Copilot (53/99), and Cursor (51/99) as the top three tools (\autoref{fig:survey_tasks_experience_tools_desired_attributes}c).
We probed into  their goals and priorities when integrating agents into their workflows, asking \studyOne participants in the interview, while priming \studyTwo participants to reflect on general development goals before specifying priorities with agents and tool choice rationale.

\mypara{Personal Productivity}
Experienced developers valued agentic tools for accelerating software development (\numStudyOne{9}, \numStudyTwo{35}).
Tool attributes affecting  productivity include their low barrier to entry (\numStudyOne{3}, \numStudyTwo{8}), integration into existing tools (\numStudyOne{3}, \numStudyTwo{11}), and quality execution of user requests (\numStudyOne{6}, \numStudyTwo{36}), along with their access to customization (\eg, user rules; \numStudyTwo{4}) and state-of-the-art models (\numStudyTwo{3}).
Participants appreciated this productivity boost at minimal cost (\numStudyOne{2}, \numStudyTwo{15}).
Finally, some started using agentic tools out of curiosity after hearing about them from colleagues and influencers (\numStudyOne{5}, \numStudyTwo{36}).

\myparaskip

\noindent
\myparaheadstyle{Maintaining Software Quality Attributes}
Experienced developers valued many quality attributes when using agents (\numStudyOne{6}, \numStudyTwo{67}). 
Though unprompted, four \studyOne participants (P2, P4, P8, P12) mentioned prioritizing modularity,  maintainability, compatibility, and correctness; two translated readability and test-driven development preferences into user rules (P6, P13; \autoref{subsec:results-strategies}).
67 \studyTwo respondents prioritized quality attributes (\autoref{fig:survey_tasks_experience_tools_desired_attributes}d), 
especially correctness
(S71: \emph{``that it wouldn't [screw] up my code''}) 
and readability (S53: \emph{``most AI agents produce messy and unnecessarily long code when not given enough instructions''}).
Conversely, 37 mentioned non-quality attributes (\eg, productivity) and 9 valued unspecified quality (\eg, S56: ``best practices'').
Overall,  software quality remains crucial for experienced developers using agents.


\mytakeaway{
\textbf{Takeaway 1:}
Experienced developers appreciate agentic tools for boosting productivity, while simultaneously valuing software quality attributes.
}

\subsection{RQ2: Strategies for Using Agents}\label{subsec:results-strategies}

\begin{table}[b]
    \centering
    \footnotesize
    \setlength{\tabcolsep}{2.5pt}
    \caption{\studyTwo ratings of agentic use.}

    \renewcommand{\arraystretch}{1.2}
    \newcommand{\qscale}[2]{\centerline{#1\,\,\,.\,\,\,.\,\,\,.\,\,\,#2}}
    \newcommand{\dist}[1]{{%
      \setlength{\fboxsep}{0pt}
      \raisebox{-0.8em}{%
        \colorbox{barcolor}{\includegraphics[height=2em]{#1}}%
      }%
    }}%

    \begin{tabular}{m{0.7\linewidth}ccc}
        \toprule
         \textbf{\textit{For this task...}} & \textbf{Mean} & \textbf{Mdn.} & \textbf{Dist.} \\
        \hline
        \textit{...how frequently did you feel like you had to modify or change the code output generated by the agent(s)?}\newline\qscale{Never (1)}{Always (5)}& 3.0 & 3 & \dist{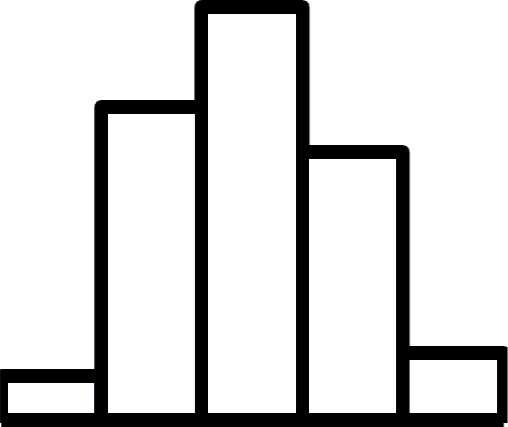} \\
        \hline
        \textit{...how suitable do you feel agent(s) are for the task?}\newline\qscale{Extremely unsuitable (1)}{Extremely suitable (6)} & 4.7 & 5 & \dist{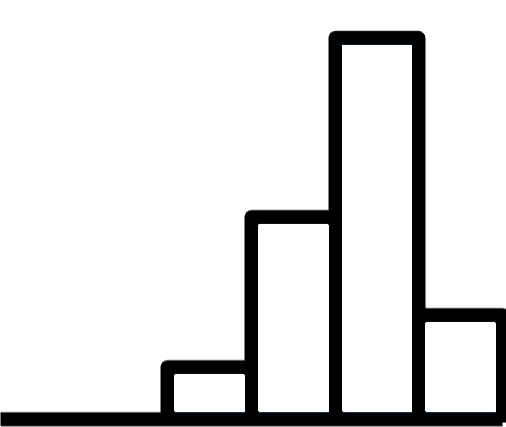} \\
        \hline
        \textit{...how much did you enjoy developing software with agents compared to without?}\newline\qscale{Extremely displeased (1)}{Extremely pleased (6)} & 5.1 & 5 & \dist{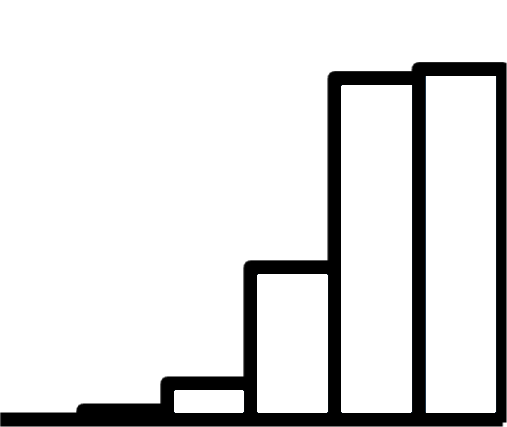} \\
        \bottomrule
    \end{tabular}
    \label{tab:categorical-survey-plots}
\end{table}

\topquote{I am a software engineer, I prompted by applying the lessons of software engineering to narrative.}{S88}


With the reported values (\autoref{subsec:results-values}), participants demonstrated several strategies  that imply a desire for \emph{controlling} agent behavior.

\mypara{Controlling Software Design and Implementation}
Although we did not explicitly ask about preference for control when working with agents, \studyTwo respondents reported modifying agent-generated code about half the time on average (\autoref{tab:categorical-survey-plots}), and 50/99 mentioned driving the architectural requirements and design.
We saw strategies from \emph{all} \studyOne participants to oversee the software design, implementation, or both (if occurred in the same task), regardless of their task domain familiarity.
When creating new features, 11 participants controlled the design (all but P5 and P13, whose tasks were entirely debugging) by creating plans themselves (P3, P4, P6, P7, P8, P9, P10, P11), revising agent-generated plan drafts (P1, P12), or asking the agent to polish human-generated plans (P2).
Maintaining full control over design planning was particularly important for ensuring stakeholder requirements (P2, P12) and keeping proprietary information intact (P2, P8).
Even when asking the agent to generate a base plan for an unfamiliar task, P1 always reviewed the plan, leveraging general software engineering expertise, to ensure they could \emph{``test [each proposed implementation] in an integrated way.''}
%
Regardless of tasks, all 13 \studyOne participants controlled the implementation at some level.
Three participants (P1, P4, P5), all working on unfamiliar tasks, specified  requirements and let the agents drive the implementation, not necessarily reviewing the agent-generated code but closely monitored the agents' decisions and program outputs in a skeptical and controlling manner, \eg, rejecting dependencies the agents would try to install (P1, P4) and using a debugger to resolve AI-generated misbehaving code (P5).
Nine (P2, P3, P6, P7, P8, P9, P10, P11, P13), all working within their domain of expertise,
carefully reviewed \emph{every} agentic change, approaching revision by prompting the agent (P3, P6, P7, P11, P13) to \emph{``keep everything in the chat context''} (P6) or manually changing the code (P2, P8, P9, P10) to \emph{``save time in the long run [by not] re-explaining everything later.''} (P10)
\autoref{tab:prompt_size_and_verification} lists their code verification strategies, such as manual UI testing, reading diffs, using debuggers, and running tests through CLI.
Unlike the others, P12 drove the implementation himself while using the agent for architectural explanation as he refactored the unfamiliar code base, but got annoyed when the agent modified a file he was actively editing without instructed: \emph{``well, that's not nice''}.

\begin{table}[b]
    \footnotesize
    \centering
    \caption{Summary of observed prompting strategies and code verification tactics. ``Plan Files''=writing/saving plans to external files, ``Context Files''=using local files to maintain agent context/memory across sessions, ``Max Size''=number of words in the largest prompt or plan, ``Max Steps''=steps in the largest prompt or plan, ``SE/P''=steps executed per prompt. SE/P can exceed the largest plan size if, \eg, a prompt references a prior plan and also requests extra steps.}
    \label{tab:prompt_size_and_verification}
    \setlength{\tabcolsep}{1.5pt}
    \rowcolors{1}{}{gray!10}
    \begin{tabular}{cccrrrr|ccc}
        \hiderowcolors
        {\textbf{\thead{\\ID}}} & {\textbf{\thead{Plan\\Files?}}} & {\textbf{\thead{Context\\Files?}}} & {\textbf{\thead{Max\\Size}}} & {\textbf{\thead{Max\\Steps}}}  & {\textbf{\thead{Max\\SE/P}}} & {\textbf{\thead{Mean\\SE/P}}} & 
        {\textbf{\thead{How\\Verified?}}} & 
        {\textbf{\thead{Reviewed\\Code Diffs?}}} & 
        {\textbf{\thead{Who\\Revised?}}}
        \global\rownum=0\relax \\
        \showrowcolors \midrule
        P1 & \checkmark & \checkmark & 917 & 70 & 6 & 2.2 & UI & & Agent \\
        P2 & \checkmark & \checkmark & 619 & 71 & 5 & 1.9 & IDE diff & \checkmark &  Human \\
        P3 & & & 225 & 5 & 3 & 1.5 & UI & \checkmark & Agent \\
        P4 & & & 84 & 4 & 4 & 1.8 & UI & & Agent \\
        P5 & & & 55 & 3 & 3 & 1.6 & UI, debugger & & Agent \\
        P6 & & & 125 & 9 & 13 & 5.0 & CLI & \checkmark & Agent  \\
        P7 & & & 91 & 2 & 2 & 1.3 & CLI & \checkmark & Agent \\
        P8 & & & 43 & 3 & 2 & 1.1 & CLI & \checkmark & Human \\
        P9 & & & 275 & 2 & 1 & 1.0 & Git diff & \checkmark & Human \\
        P10 & & & 288 & 7 & 7 & 3.0 & UI, dev tools$^*$ & \checkmark & Human \\
        P11 & \checkmark & \checkmark & 189 & 11 & 12 & 3.5 & CLI & \checkmark & Agent \\
        P12 & & & 48 & 3 & 5 & 2.3 & N/A$^\dagger$ & N/A$^\dagger$ & Human \\
        P13 & & & 558 & 3 & 3 & 1.4 & CLI & \checkmark & Agent \\
        \hiderowcolors \midrule
        \multicolumn{3}{r}{\textit{Min}} & \textit{43} & \textit{2} & \textit{1} & \textit{1.0} & \\
        \multicolumn{3}{r}{\textit{Median}} & \textit{189} & \textit{4} & \textit{4} & \textit{1.8} & \multicolumn{3}{l}{\quad\textit{$^*$browser dev tools}}\\
        \multicolumn{3}{r}{\textit{Max}} & \textit{917} & \textit{71} & \textit{13} & \textit{5.0} & \multicolumn{3}{l}{\quad\textit{$^\dagger$no agent-generated code (P12)} }\\
        \bottomrule

    \end{tabular}
\end{table}

\colorlet{tagFeature}{magenta}
\colorlet{tagDomain}{teal}
\colorlet{tagInteraction}{green!60!black}
\colorlet{tagUI}{blue}
\definecolor{tagTech}{rgb}{0.49,0.14,0.81}
\colorlet{tagFile}{orange}
\colorlet{tagStep}{red!80!black}

\begin{table}[t]
\caption{Top 10 types of context included in prompts executed by each participant. Colors correspond to the \autoref{fig:example_prompt} example.}%
\label{tab:prompt_context}%
\centering%
\footnotesize%
\setlength{\tabcolsep}{1.45pt}
\renewcommand{\arraystretch}{1.1}
\newcommand{\newcm}[0]{\checkmark}
\newcommand{\nocm}[0]{}
\rowcolors{1}{}{gray!10}%
\begin{tabular}{l|ccccccccccccc}%
\hiderowcolors%
 & \textbf{P1} & \textbf{P2} & \textbf{P3} & \textbf{P4} & \textbf{P5} & \textbf{P6} & \textbf{P7} & \textbf{P8} & \textbf{P9} & \textbf{P10} & \textbf{P11} & \textbf{P12} & \textbf{P13} \global\rownum=0\relax \\
\showrowcolors \midrule
\textcolor{tagTech}{Technical Term} & \checkmark &  & \checkmark & \checkmark & \checkmark & \checkmark & \checkmark & \checkmark & \checkmark & \newcm & \checkmark & \checkmark & \newcm \\
\textcolor{tagDomain}{Domain Object} &  & \checkmark & \newcm & \checkmark & \checkmark & \checkmark & \newcm &  &  & \checkmark & \newcm & \newcm & \newcm \\
\textcolor{tagFeature}{New Feature or Requirement} & \checkmark & \checkmark & \newcm & \newcm & \newcm & \checkmark & \newcm & \checkmark &  & \checkmark & \newcm &  &  \\
Reference to Input File & \checkmark & \checkmark & \checkmark &  &  &  & \checkmark & \checkmark & \checkmark & \checkmark & \checkmark & \checkmark & \checkmark \\
Specific Library or API & \checkmark & \checkmark &  & \checkmark &  & \checkmark & \checkmark & \checkmark & \checkmark & \checkmark &  & \checkmark & \newcm \\
\textcolor{tagInteraction}{Interaction} & \checkmark & \checkmark & \checkmark & \checkmark & \newcm &  &  &  & \checkmark & \checkmark & \newcm &  &  \\
Purpose of Feature & \checkmark & \checkmark &  & \newcm & \newcm & \checkmark &  & \nocm &  & \nocm & \newcm & \checkmark & \newcm \\
\textcolor{tagUI}{UI or Design Term} & \checkmark & \checkmark & \checkmark & \checkmark & \checkmark &  &  &  & \checkmark & \checkmark &  &  &  \\
\textcolor{tagFile}{Reference to Output File} & \checkmark & \checkmark & \checkmark & \checkmark &  & \newcm &  & \newcm &  &  &  & \newcm &  \\
\textcolor{tagStep}{Reference to Step in Plan} & \checkmark & \checkmark & \checkmark & \checkmark & \checkmark &  &  &  &  &  &  &  &  \\
\bottomrule
\end{tabular}
\end{table}

\mypara{Controlling Agent Behavior through Prompting Strategies}
We identified several prompting strategies experienced developers employed to improve agentic outputs.
Most agreed prompts require clear context and explicit instructions (\numStudyOne{12}, \numStudyTwo{43}); \eg, S59 described providing filenames, variables, and error messages to ensure \emph{``small and specific modifications.''}
To understand this context requirement, in addition to reviewing strategies reported in \studyTwo 
(see supplement),
we analyzed \studyOne prompts, finding that they incorporated various context types (\autoref{tab:prompt_context}).
For instance, one P3's prompt (\autoref{fig:example_prompt}) used 7/10 top types
and restricted the agent to \emph{``do just step 1''} to avoid premature execution of the complex prompt.
We further measured prompt and plan sizes to assess prompt complexity, listing the largest plan or prompt per participant in \autoref{tab:prompt_size_and_verification}.
While the largest plans from P1, P2, and P11 exceeded 10 steps (P1 and P2's plans built with agents exceeded 70), they were executed in small chunks to maintain control.
Specifically, P1 and P2 never executed over 6 steps simultaneously, and participants assigned a median of 1.8 steps per prompt.

Participants also employed mechanisms like screenshots (\numStudyOne{3}), file references (\numStudyOne{5}), examples (\numStudyOne{1}, \numStudyTwo{5}), step-by-step thinking (\numStudyOne{1}, \numStudyTwo{3}), and external information via Model Context Protocols (\numStudyOne{1}, \numStudyTwo{5}), treating AI as \emph{``a smart kid that knows nothing about you and the outside world''} (S75).
Some used external agents to refine prompts (\numStudyTwo{3}), or applied user rules (\numStudyOne{3}, \numStudyTwo{18}) to enforce project specifications (\numStudyOne{2}, \numStudyTwo{5}), provide general guidelines (\numStudyTwo{8}), or correct agent behaviors (\numStudyOne{2}), effectively \emph{``applying the lessons of software engineering to narrative''} (S88).

Context length preferences varied: some kept long chats to retain context (\numStudyOne{2}, \numStudyTwo{29}) or iterate (\numStudyOne{2}, \numStudyTwo{8}) on errors (\numStudyTwo{2}); others kept contexts short (\numStudyOne{5}, \numStudyTwo{3}) to focus on small tasks (\numStudyOne{2}, \numStudyTwo{2}) and minimize waiting (\numStudyOne{2}, \numStudyTwo{1}).
Ultimately, prompting remained a self-taught \emph{``trial and error''} process (P12)---developers continually mixed 
strategies (\numStudyTwo{13}).

\begin{figure}
\newcommand{\annotatedtext}[3]{%
  \textcolor{#1}{#2}~%
  \raisebox{1pt}{\tcbox[
    on line, 
    arc=2.65pt, 
    colback=#1!10!white, 
    colframe=#1, 
    boxrule=0.5pt, 
    boxsep=0pt, 
    left=2pt, 
    right=2pt, 
    top=0.8pt, 
    bottom=0.2pt
  ]{\textcolor{#1}{\fontsize{4.25pt}{4.25pt}\selectfont\strut #3}}}%
}
\promptbox{{\sffamily\fontsize{7pt}{8.1pt}\selectfont\textls[-20]{%
\begin{justify}%
\vspace{-0.65em}
Let's add a \annotatedtext{tagFeature}{new feature that lets a user score}{New Feature or Requirement} \annotatedtext{tagDomain}{annotations}{Domain Object} from 1 to 4.\\[0.5em]
In the UI, a user can score an annotation by \annotatedtext{tagInteraction}{hovering over that annotation}{Interaction} and pressing the key 1, 2, 3, or 4. When this happens, the score should be updated in the frontend dataset. The UI should display annotation scores with a \annotatedtext{tagUI}{small indicator in a bubble on the bottom-left of}{UI or Design Term} each annotation highlight. For now, we will ignore the annotations panel and focus entirely on the main content panels. As always, when the user presses save, the annotations data should be persisted to the backend. This includes the new annotation scores.\\[0.5em]
Here are the steps to follow to implement this:
\begin{enumerate}[nosep, label=\arabic*., align=left, labelwidth=1em, labelsep=0pt, leftmargin=!]
\item Update the annotation data type. Add an optional ``quality'' field to both the \annotatedtext{tagTech}{TextLabel}{Technical Term} and the \annotatedtext{tagTech}{TextMapping}{Technical Term} types.
\item Add new \annotatedtext{tagTech}{keybindings}{Technical Term} to the \annotatedtext{tagUI}{frontend that listen for the keys 1, 2, 3, and 4.}{UI or Design Term} Take a look at how \annotatedtext{tagTech}{keyEventUtils}{Technical Term} is set up and properly update the \annotatedtext{tagTech}{handleApplicationLevelHotkeys}{Technical Term} function to support the new listener. Implement the listener callback within \annotatedtext{tagFile}{App.tsx}{Reference to Output File}.
\item Update the UI rendering to draw the score indicators, when a score exists, on each annotation highlight in the main content panels.
\end{enumerate}\vspace{0.5em}
We will do these steps one at a time. \annotatedtext{tagStep}{Please do just step 1 now.}{Reference to Step in Plan}%
\end{justify}%
}}}
    \caption{A prompt from P3, with seven of the ten context types of~\protect\autoref{tab:prompt_context}. Select examples indicated by colored tags.}
    \label{fig:example_prompt}
\end{figure}

\mypara{Controlling Agent Behavior through Software Engineering Skills}
Beyond prompting, participants leveraged traditional software engineering practices to control agents.
\emph{All} \studyOne participants oversaw agent performance during planning and code generation, using techniques like reading diffs (P2, P6, P10, S103) and testing (P3, P4, P8, P10, P12, P13); S97 would even \emph{``read every line of code before accepting.''}
Notably, five \studyOne participants tested code more systematically when using agents, with P6 having agents generate tests for every change to reinforce test-driven development.
However, testing remained manual for UI interactions (P4, P10), databases (P8), and remote applications (P10).
Participants also proactively intervened by monitoring the agent's thinking process (P4, P5, P6, P10), decomposing complex tasks into execution steps (P4, P7, P10), revising code to demonstrate good practices \eg, readability (P2, P6), and overriding behavior using human expertise (P1, P2).
When debugging agent-generated errors, some offloaded analysis to agents (P4, P10), while others like S15 and P12 refused to \emph{``waste time''} with them at the system level.
Finally, participants leveraged traditional version control to both ease rollbacks (P4, P10, S73, S85) and merge work across multiple agents (\numStudyOne{2}, \numStudyTwo{4}). 
The use of multiple agents, though manually configured (\numStudyTwo{6}), seemed common (\numStudyOne{4}, \numStudyTwo{31}) for 
parallelizing tasks (\numStudyOne{3}, \numStudyTwo{17}) or combining agent abilities (\numStudyOne{1}, \numStudyTwo{10}).

Ultimately, participants cited their existing domain expertise as critical for effective agent use (\numStudyOne{11}, \numStudyTwo{65}).
This required not just awareness of agent capabilities (P1, P5), but strong code comprehension and debugging skills (P2, P3, P6, P7, P10, P12), and the ability to translate product specs into accurate implementation plans before generation (\numStudyOne{3}, \numStudyTwo{13}).
As S2 noted, it is vital to \emph{``first [understand] coding and then [use] AI.''}

\vspace{1.5em} 
\mytakeaway{
\textbf{Takeaway 2:}
With agents, experienced developers \textit{control}  software design and implementation
by prompting and planning with \textit{clear context and explicit instructions}, and letting agents work on only a few tasks at a time.
They also leverage established SWE best practices, \eg, validation and version control, and expertise to 
incorporate agentic changes with supervision.
}

\subsection{RQ3: Agentic Task Suitability}\label{subsec:results-suitability}


\topquote{There's almost nothing I won't use these for. Even for one line items I feel confident in implementing myself, I like walking through it with the agent. It's a second set of eyes...}{S8}

\definecolor{scolor}{rgb}{0.1,0.1,0.95}
\definecolor{ucolorlight}{rgb}{1,0,0.15}
\definecolor{ucolor}{rgb}{0.87,0,0.1305}
\definecolor{ccolor}{rgb}{0.5,0.5,0.3}
\newcommand{\SU}[2]{\textcolor{scolor}{#1}\raisebox{0.05em}{\textcolor{gray}{:}}\textcolor{ucolor}{#2}}
\newcommand{\taskname}[2]{{\sffamily\textcolor{#1}{#2}}}
\newcommand{\taskSU}[3]{%
  \begingroup
    \count0=\numexpr 10*#2\relax
    \count1=\numexpr 25*#3\relax
    \ifnum\count0>\count1
      \taskname{scolor}{#1} (\SU{#2}{#3})%
    \else\ifnum\count0=\count1
      \taskname{scolor}{#1} (\SU{#2}{#3})%
    \else
      \count0=\numexpr 10*#3\relax
      \count1=\numexpr 25*#2\relax
      \ifnum\count0>\count1
        \taskname{ucolor}{#1} (\SU{#2}{#3})%
      \else\ifnum\count0=\count1
        \taskname{ucolor}{#1} (\SU{#2}{#3})%
      \else
        \taskname{ccolor}{#1} (\SU{#2}{#3})%
      \fi\fi
    \fi\fi
  \endgroup
}

On average, \studyTwo respondents rated the agents' suitability for their stated task as 4.73/6 (6=``Extremely suitable'', see \autoref{tab:categorical-survey-plots}). 
%
We coded the suitability of their stated task and other tasks mentioned in the survey.
\autoref{tab:suitability} lists the 59 fine-grained task codes mentioned in at least 5 surveys, grouped by the respondents' general sentiment about agent suitability: task codes with at least a 2.5:1 ratio of suitable to unsuitable responses are categorized as ``Maybe~suitable'', those with less than 1:2.5 as ``Maybe~unsuitable'', and the remainder as ``Controversial''. 
Per the reflexive nature of this analysis~\cite{braunUsingThematicAnalysis2006}, the thresholds were selected based on the authors' judgment.
Below we discuss the major themes that \autoref{tab:suitability} suggests.
We write task codes as \taskname{ccolor}{task name} (\SU{S}{U}) where \textcolor{scolor}{S} and \textcolor{ucolor}{U} are the number of  responses indicating \textcolor{scolor}{suitability} and \textcolor{ucolor}{unsuitability}; 
see the supplement for a full table with respondent quotes.

\begin{table}[t]
\caption{
Number of survey responses mentioning a task as \textcolor{scolor}{suitable (S)} or \textcolor{ucolor}{unsuitable (U)} for AI agents.
}
\label{tab:suitability}
{
\rowcolors{1}{}{gray!10}
\fontsize{7.25pt}{7.75pt}\selectfont 
\centering
\setlength{\tabcolsep}{1.7pt} 
\newcommand{\barscale}{0.8}
\newcommand{\SUbar}[2]{\pgfmathparse{\barscale*23-\barscale*#2}\hspace{\pgfmathresult pt}\raisebox{-0.31em}{\tikz \fill[ucolorlight] (0,0) rectangle (\barscale*#2pt, 1.1em);\tikz \fill[scolor] (0,0) rectangle (\barscale*#1pt, 1.1em);}}
\centerline{\begin{tabular}{c r c l} 
    \hiderowcolors
    \toprule
     & 
    \textbf{\relsize{1}Task or Use Case} &
    \textbf{\relsize{1}\SU{S}{U}} &
\hspace{4pt}\textbf{\relsize{1}\textcolor{ucolor}{\clipbox{0 0 3pt 0}{$\leftarrow$}\,U}} \textbf{\relsize{1}\textcolor{scolor}{S\,\clipbox{3pt 0 0 0}{$\rightarrow$}}}
     \\
    \hiderowcolors
    \midrule
    \showrowcolors
\cellcolor{white} & Accelerating productivity & \SU{35}{2} & \SUbar{35}{2}\\
\cellcolor{white} & Small/simple/straightforward tasks & \SU{33}{1} & \SUbar{33}{1}\\
\cellcolor{white} & Following well-defined plans & \SU{28}{2} & \SUbar{28}{2}\\
\cellcolor{white} & Generating new code & \SU{27}{2} & \SUbar{27}{2}\\
\cellcolor{white} & Tedious/repetitive tasks & \SU{26}{0} & \SUbar{26}{0}\\
\cellcolor{white} & Scaffolding or boilerplate & \SU{25}{0} & \SUbar{25}{0}\\
\cellcolor{white} & Writing tests & \SU{19}{2} & \SUbar{19}{2}\\
\cellcolor{white} & Refactoring (general) or code improvement & \SU{18}{3} & \SUbar{18}{3}\\
\cellcolor{white} & All or most all tasks & \SU{20}{0} & \SUbar{20}{0}\\
\cellcolor{white} & Writing/updating documentation & \SU{20}{0} & \SUbar{20}{0}\\
\cellcolor{white} & Backend development & \SU{14}{1} & \SUbar{14}{1}\\
\cellcolor{white} & Debugging (simple) or simple fixes & \SU{12}{3} & \SUbar{12}{3}\\
\cellcolor{white} & Refactoring (simple) or modifying code or cleanup & \SU{14}{0} & \SUbar{14}{0}\\
\cellcolor{white} & Starting or setting up a new project & \SU{10}{4} & \SUbar{10}{4}\\
\cellcolor{white} & Frontend development & \SU{10}{3} & \SUbar{10}{3}\\
\cellcolor{white} & Prototyping or small projects & \SU{12}{0} & \SUbar{12}{0}\\
\cellcolor{white} & Explaining/analyzing code/APIs/errors & \SU{10}{1} & \SUbar{10}{1}\\
\cellcolor{white} & Exploring alternatives or experimenting & \SU{7}{0} & \SUbar{7}{0}\\
\cellcolor{white} & Helping the developer learn & \SU{6}{1} & \SUbar{6}{1}\\
\cellcolor{white} & Looking up knowledge for the developer & \SU{6}{1} & \SUbar{6}{1}\\
\cellcolor{white} & Data processing/analysis & \SU{5}{2} & \SUbar{5}{2}\\
\cellcolor{white} & Collaboratively talking out problems & \SU{6}{0} & \SUbar{6}{0}\\
\cellcolor{white} & Writing helper functions & \SU{6}{0} & \SUbar{6}{0}\\
\cellcolor{white} & Testing & \SU{5}{0} & \SUbar{5}{0}\\
\cellcolor{white} & LLM following a knowledge base & \SU{4}{1} & \SUbar{4}{1}\\
    \multirow{-26}{*}{\rotatebox[origin=c]{90}{\textbf{\relsize{1}Maybe suitable for...}}}
\cellcolor{white} & Using major/common frameworks/libraries & \SU{4}{1} & \SUbar{4}{1}\\
    \hiderowcolors
    \midrule
    \showrowcolors
\cellcolor{white} & High level plans (\eg project or architecture design) & \SU{13}{23} & \SUbar{13}{23}\\
\cellcolor{white} & Debugging (general) & \SU{12}{8} & \SUbar{12}{8}\\
\cellcolor{white} & Brainstorming or conceptualization & \SU{11}{5} & \SUbar{11}{5}\\
\cellcolor{white} & UI/HTML/CSS & \SU{9}{4} & \SUbar{9}{4}\\
\cellcolor{white} & Understanding project architecture & \SU{6}{7} & \SUbar{6}{7}\\
\cellcolor{white} & Long context sessions or between-task consistency & \SU{4}{8} & \SUbar{4}{8}\\
\cellcolor{white} & Code review & \SU{7}{4} & \SUbar{7}{4}\\
\cellcolor{white} & Writing feature-level implementation plans & \SU{6}{3} & \SUbar{6}{3}\\
\cellcolor{white} & Creative vision or creative tasks & \SU{4}{4} & \SUbar{4}{4}\\
\cellcolor{white} & Understanding task context & \SU{3}{4} & \SUbar{3}{4}\\
\cellcolor{white} & Domains/libraries unfamiliar to the developer & \SU{4}{2} & \SUbar{4}{2}\\
\cellcolor{white} & Creating configurations & \SU{3}{3} & \SUbar{3}{3}\\
\cellcolor{white} & Version control & \SU{2}{4} & \SUbar{2}{4}\\
    \multirow{-14}{*}{\rotatebox[origin=c]{90}{\textbf{\relsize{1}Controversial}}}
\cellcolor{white} & Handling many files at once & \SU{2}{3} & \SUbar{2}{3}\\
    \hiderowcolors
    \midrule
    \showrowcolors
\cellcolor{white} & One-shotting code without modification/verification & \SU{5}{23} & \SUbar{5}{23}\\
\cellcolor{white} & Integrating with existing code; legacy code & \SU{3}{17} & \SUbar{3}{17}\\
\cellcolor{white} & Complex tasks (not necessarily big) & \SU{3}{16} & \SUbar{3}{16}\\
\cellcolor{white} & Business logic or tasks requiring domain knowledge & \SU{2}{15} & \SUbar{2}{15}\\
\cellcolor{white} & Writing performant code & \SU{3}{9} & \SUbar{3}{9}\\
\cellcolor{white} & Replacing human expertise or decision making & \SU{0}{12} & \SUbar{0}{12}\\
\cellcolor{white} & Handling CI/deployment infrastructure & \SU{3}{8} & \SUbar{3}{8}\\
\cellcolor{white} & Handling details & \SU{2}{7} & \SUbar{2}{7}\\
\cellcolor{white} & Big tasks (not necessarily complex) & \SU{1}{7} & \SUbar{1}{7}\\
\cellcolor{white} & High-stakes or privacy-sensitive tasks & \SU{0}{8} & \SUbar{0}{8}\\
\cellcolor{white} & Security-critical code & \SU{2}{5} & \SUbar{2}{5}\\
\cellcolor{white} & Vague/unclear/open-ended tasks & \SU{0}{7} & \SUbar{0}{7}\\
\cellcolor{white} & Using uncommon or private libraries/languages & \SU{1}{5} & \SUbar{1}{5}\\
\cellcolor{white} & Complex logic & \SU{0}{6} & \SUbar{0}{6}\\
\cellcolor{white} & Choosing technologies/frameworks/libraries & \SU{1}{4} & \SUbar{1}{4}\\
\cellcolor{white} & Refactoring (complex) or improving architecture & \SU{1}{4} & \SUbar{1}{4}\\
\cellcolor{white} & Tasks highly familiar to the developer & \SU{1}{4} & \SUbar{1}{4}\\
\cellcolor{white} & Database design/management & \SU{0}{5} & \SUbar{0}{5}\\
    \multirow{-19}{*}{\rotatebox[origin=c]{90}{\textbf{\relsize{1}Maybe unsuitable for...}}}
\cellcolor{white} & New API versions after LLM's knowledge cutoff & \SU{0}{5} & \SUbar{0}{5}\\
    \bottomrule
\end{tabular}}

}

\end{table}

\mypara{AI Accelerates Straightforward, Repetitive, Scaffolding Tasks}
A large number of survey respondents reported they found agents helpful for \taskSU{accelerating productivity}{35}{2},
with many reporting preferences for agents  on \taskSU{all or most all tasks}{20}{0}, \eg \emph{``AI assistants have improved [my] experience everywhere in general''} (S55).
Unsurprisingly, respondents mentioned agents were suitable for \taskSU{generating new code}{27}{2}.
More specifically, for \taskSU{small/simple/straightforward tasks}{33}{1} and \taskSU{tedious/repetitive tasks}{26}{0} agents \emph{``removed the mundane''} (S52), although suitability for simple tasks does not extend to \emph{complex} tasks (more below). 
Agents also help write \taskSU{scaffolding or boilerplate}{25}{0}, \eg \emph{``boilerplate code that we wish we didn't have to write in the first place''} (S7). 
Agents can also assist with \taskSU{starting or setting up a new project}{10}{4}, like
	\emph{``building anything prototype from zero until [it becomes a] small app''} (S98). 

\mypara{Suitable for Following Well-Defined Plans}
AI models in 2025 may be able to follow multi-step instructions, as S25 noted: \emph{``A few months ago, you had to break down tasks to be pretty small in order to get quality output. But the new models are surprisingly good at developing even larger tasks.''}
Consequently, lots of respondents reported the agents' success at \taskSU{following well-defined plans}{28}{2},
although many emphasized the need for clear prompting, 
\eg \emph{``AI can really go off the rails and you spend more time putting it back than you gained, if you don't stay pretty specific with most things''} (S7).
The supplement
lists more quotes that emphasize the need for clarity in successfully prompting the agents.
Conversely, \emph{``if you don't lead [or] plan then you will [get] stuck''} (S89): several respondents mentioned the agents performed poorly with \taskSU{vague/unclear/open-ended tasks}{0}{7}, and a few mentioned trouble with \taskSU{long context sessions or between-task consistency}{4}{8}, or the agent \taskSU{understanding task context}{3}{4}, \eg S87 noted the agent \emph{``focused too much on specific details of the code without considering the context, which resulted in many more details being changed than required''}. 


\mypara{Suitable for General SWE Tasks: Writing Tests, General Refactoring, Documentation, Simple Debugging}
Respondents mentioned many non-implementation software engineering (SWE) activities that suit the agents, including
\taskSU{writing tests}{19}{2}, \eg 
\emph{``they are surprisingly good at writing general tests for already existing code''} (S65),
\taskname{scolor}{refactoring (simple) or modifying code or cleanup} (\SU{14}{0}), and \taskname{scolor}{refactoring (general) or code improvement} (\SU{18}{3}); 
this aligns with recent findings that 39\% agent-generated commits to open-source Java projects contained refactoring~\cite{AgenticRefactoring2025}.
Note we subdivided refactoring and debugging into simple/general/complex tasks codes.
Many also used agents for \taskSU{writing/updating documentation}{20}{0}, although not elaborating on their specific writing process; this accords with \citet{AIDevCorpus}'s finding that Codex-generated documentation PRs saw higher GitHub acceptance rates than human PRs.
Respondents also found agents suitable for  \taskSU{debugging (simple) or simple fixes}{12}{3}, \eg \emph{``it was highly effective at fixing a bug when I could point it to a specific failing test''} (S73), and for \taskSU{explaining/analyzing code/APIs/errors}{10}{1} with S31 noting \emph{``agents are good at breaking down confusing stack traces and suggesting possible causes''}.

Beyond \emph{simple} debugging (\SU{14}{0}) or \emph{general} refactoring (\SU{18}{3}), opinions were mixed.
``Complex'' debugging was not mentioned enough to report, but \taskSU{Debugging (general)}{12}{8} was controversial. 
For example, S57 reported debugging with LLMs \emph{``often caus[ed] more problems than they initially found''}, whereas \newS{103} preferred agents for \emph{``debugging (but they can get stuck in debug loops)''}. 
%
Although suitable for \emph{general} refactoring (\SU{18}{3}), agents were viewed negatively for \taskSU{refactoring (complex) or improving architecture}{1}{4}:
\eg S79 avoided agents for \emph{``complicated, many-step, boundary-crossing changes across codebases that have to be coordinated''}.
As task complexity increases, agents may become less suitable.

\mypara{Suitable for Backend and Some Frontend Work}
Both backend and frontend developers reported the agents' suitability for their work (\SU{14}{1} and \SU{10}{3} for \taskname{scolor}{backend} and \taskname{scolor}{frontend development} respectively).
%
Within frontend work, however, specific mentions of \taskSU{UI/HTML/CSS}{9}{4} were positive but not universally so:
while S95 said \emph{``the UI part of my work (SwiftUI) was done well''} and S26 reported \emph{``[agents] help me save time to do UI layout''}, S23 disliked agents for \emph{``fine tuning CSS/HTML''}, and S24 said \emph{``there were often little things in the UI that were not right that were time consuming to fix''}.
Likewise, P1 in \studyOne said they tried \emph{``I don't know, 30 times''} to make an animated web banner but the AI \emph{``just wouldn't achieve what I was looking for''}.
These experiences suggest that agents may be helpful with UI construction but not for controlling specific visual details.

\mypara{Suitable for Prototyping and Experimenting}
Respondents found agents suitable for \taskSU{prototyping or small projects}{12}{0}: 
\eg \emph{``building a framework/prototype of the solution [...] around 2k LoC is straight forward with an LLM and can even be vibe coded efficiently''} (S15).
Some respondents mentioned \taskSU{exploring alternatives or experimenting}{7}{0}, \eg using agents for \emph{``throwaway demos/experiments''} (S24).

\mypara{Controversial for Plan Development}
The most interesting disagreement is whether agents are suitable for \taskSU{high level plans (\eg project or architecture design)}{13}{23}.
Most respondents who mentioned using agents for high-level planning deemed them unsuitable,
believing that agents are unsuitable for  \taskSU{business logic or tasks requiring domain knowledge}{2}{15}
and cannot \taskSU{replace human expertise or decision making}{0}{12}, detailed below.
While opinions were mixed about agents for \taskSU{understanding project architecture}{6}{7}, some considered them unsuitable for \taskSU{choosing technologies/frameworks/libraries}{1}{4}. 
For example, S68 will not use agents for \emph{``system design! I never trust LLMs for systematic issues''} and
with naive delegation to the agent \emph{``a lot of [agents'] architectural decisions [...] can come to bite back''} (S15).
%
While no respondent said agents could replace human judgment, some still used agents for high-level planning (\SU{13}{23})
perhaps due to assistance short of full delegation: agents can assist with \taskSU{brainstorming or conceptualization}{11}{5} or \taskSU{collaboratively talking out problems}{6}{0}, \emph{``like full time rubber ducking''} (S7) or \emph{``asking a co-developer for questions and insights without the fear of being judged''} (S58).
P13 noted, \emph{``I explicitly asked [the agent] to challenge me and [...] double check what I was proposing.''}
Similarly, S35 preferred \emph{not} to use agents for \emph{``high-level planning, [though they] help me avoid blind spots. [It's] a lot of back-and-forth, so I don't leave agents to work on that without my input.''}
P12 called agents a \emph{``lifesaver''} because he would use them to draw the architecture of the system to help him visually reason about how new features would integrate.
These experiences suggest that agents can be a collaborator, rather than an autonomous driver, for system and architecture design.

At a less high level, a handful of respondents used agents for \taskSU{writing feature-level implementation plans}{6}{3} like \emph{``PRD generation, breaking down [tasks]''} (S55).
But three respondents were negative about it, \eg \newS{100} said \emph{``planning...went off the rails frequently''}.
%
%

\mypara{Unsuitable for Business Logic, Tasks Requiring Domain Knowledge, and Human Decision Making}
Several reported that agents are unsuitable for \taskSU{business logic or tasks requiring domain knowledge}{2}{15}
like \emph{``business logic that requires a lot of context''} (S95).
Similarly, 12 respondents deemed agents unsuitable for \taskname{ucolor}{replacing human expertise or decision making} (\SU{0}{12}),
\eg
S59 emphasized \emph{``decision making on implementing an agent generated plan/option/modification lies with me - I take my time thinking about its implication before I give a thumbs up''.}
Importantly, \textbf{no respondent indicated that agents are suitable for completely autonomous operation}, instead the more common sentiment was that of S83: \emph{``I do everything with *assistance* but never let the agent be completely autonomous---I am always reading the output and steering.''}

\mypara{Unsuitable for Perfect Code Generation}
Human oversight is still needed because LLMs may  be unsuitable for \taskSU{one-shotting code without modification/verification}{5}{23}
and \taskSU{handling details}{2}{7}.
That is, generated code needs revision, verification, or cleanup: \emph{``Almost everything it performs great on. It just does not one shot things''} (S49)
and the agent can suffer from \emph{``over engineering, not as detail-aware, [it] keep repeating itself''} (S94).
8 \studyOne participants said agents might do more than asked, \eg create extra files, rewrite entire code sections, or install unneeded packages, a trouble found in other studies~\cite{Waseem2025VibeCoding,BuildingSoftwareByRollingTheDice}. Put bluntly, P7 said, \textit{``I do not quite enjoy the fact that it will just like start changing and changing stuff without me actually understanding what's going on''}.
An oft-mentioned problem was \taskSU{integrating with existing code or handling legacy code}{3}{17}, \eg the agent \emph{``often wrote custom code for pre-existing functions in core libraries''} (S80)
and \emph{``if I want to explore or understand other legacy codebases, my experience is not so good''} (S92). 
%
%
Although agents cannot one-shot code, a few respondents (and P5, P6, P9) used agents for \taskSU{code review}{7}{4},
although others worried, 
\eg S13 reviewed themself due to accountability: 
\emph{``I am responsible for all code and deliverables, so I will independently review the content generated by the agent''}.

\mypara{Unsuitable for Complex Tasks}
In contrast to simple and repetitive tasks, agents become less suitable as task complexity increases.
Respondents said agents struggled at \taskSU{complex tasks (not necessarily big)}{3}{16}, specifically at \taskSU{complex logic}{0}{6} and, as noted previously, at \taskSU{refactoring (complex) or improving architecture}{1}{4}.
For example, S4 \emph{``never finished the task''} because \emph{``my use case is complicated and [the agent] performed poorly throughout most of it''}
and S16 found \emph{``if the task is complex, like splitting lots of changes in a work directory to a bunch of smaller commits, it can [get] stuck and do bad things''}.
Several reported agent unsuitability for \taskSU{big tasks (not necessarily complex)}{1}{7}, \eg \emph{``agent often misses following the guidelines and at times marks incorrect changes''} (S55).
In contrast, P3 from \studyOne said he completed a migration to React in three days with agents that
\emph{``would have taken probably two weeks of my own time''} and 
\emph{``I don't think we've caught a single bug''} since its completion.
We speculate this task may have worked because React is a common library, and prior code scaffolded the migration.

\mypara{Misc Unsuitable: Performance-Critical, Deployment, High-Stakes, and Security-Critical Situations}
Respondents considered some other tasks unsuitable for agents,
\eg, \taskSU{writing performant code}{3}{9} like \emph{``high performance structural design''} (S43),
or \taskSU{handling CI/deployment infrastructure}{3}{8}---they generally preferred doing so themselves.
Respondents also avoided agents for \taskSU{high-stakes or privacy-sensitive tasks}{0}{8} or \taskSU{security-critical code}{2}{5} 
given little room for error, \eg
S64 avoids agents for 
\emph{``implementing critical components that need to be performant or secure''}.

\mytakeaway{
\textbf{Takeaway 3a:}
Experienced developers find agents \textit{\textcolor{scolor}{suitable} for accelerating straightforward, repetitive, and scaffolding tasks if prompted with well-defined plans,}
Beyond writing new code and prototyping, these tasks include writing tests, documenting, general refactoring and simple debugging.

\vspace{0.5em}
\textbf{Takeaway 3b:}
\textit{As task complexity increases, agent suitability decreases.}
Agents may be \textit{\textcolor{ucolor}{unsuitable} for tasks needing domain knowledge} such as business logic, and \emph{no} respondent said agents could replace human decision making.

\vspace{0.5em}
\textbf{Takeaway 3c:}
Experienced developers \textit{\textcolor{ccolor}{disagree} about using agents for software planning and design}. Some avoided agents out of concern over the importance of design, while others embraced back-and-forth design with an AI.
}

\subsection{RQ4: Sentiments Towards Agents}\label{subsec:results-sentiments}

\topquote{It felt like driving a F1 car. While it also felt like getting stuck in traffic jam a lot, I still felt optimistic about it.}{S24}

On average, \studyTwo participants estimated 5.11/6 enjoyment of coding with agents compared to without (\autoref{tab:categorical-survey-plots}), a positivity shared by all \studyOne participants.
Below we detail reasons for the overall enjoyment, emphasizing necessary human oversight.
\mypara{Happiness, Fun, Curiosity, Relief}
Participants reported positive feelings after working with agents (P3, P7, P8): \emph{``If it's a simple task, it most likely will get correct, and I will have something neat and cool [to] use directly, [and] I feel good''} (P7).
Some found coding fun again (P3, S8)---feeling \emph{``like rediscovering computers again for the first time''} (S8)---and expressed more curiosity about the agents' capabilities (\numStudyOne{2}, \numStudyTwo{2}).
Compared to manual coding, agents reduced stress (P10, S59) by shifting focus towards \emph{``how to structure my prompts''} (S59) and bringing developers closer to the software they wanted to create (\numStudyOne{5}).
Consequently, regardless of whether their study tasks fell within (P2, P3, P6, P8) or outside (P1, P4, P5) their typical expertise, participants felt hopeful that agents would empower them to build software in unfamiliar domains.


\mypara{Agents for Collaboration, with Humans in the Loop}
Developers used agents to clarify their thought processes (\numStudyOne{4}, \numStudyTwo{12}), with P12 employing them as a \emph{``rubber ducky''} to talk through coding steps.
More broadly, agents served as collaborative brainstorming partners for abstract, big-picture thinking rather than just code generation (\numStudyOne{5}, \numStudyTwo{34}).

Notably, experienced developers preferred maintaining \emph{control} when brainstorming with agents, driving the conversation and thinking themselves and treating agents merely as inspiration (\numStudyOne{5}, \numStudyTwo{12}).
P2 doubted if we are ready to \emph{``hand off the responsibility of software engineering [...] to an AI,''} and others (\numStudyOne{6}) stressed the necessity of human codebase understanding for production reviews.
This directly impacts maintainability: \emph{``[agents] can [be] a crutch to [...] have people who don't know what they're doing generate something that seemingly works, but it is kind of a nightmare to maintain anything.''} (P4)
Overall, strong software engineering foundations remain critical for controlling, evaluating, and correcting the agent (\numStudyOne{5}, \numStudyTwo{5}).
Ultimately, \emph{``you still do have to think about it yourself [...] and evaluate if what the LLM is telling us is a good approach.''} (P8)
%

\mytakeaway{
\textbf{Takeaway 4:}
Experienced developers enjoy working \emph{with} agents as source of collaboration---with the human in the loop---rather than delegating work to the agents completely.
}

\section{Discussion}
\label{sec:discussion}


Our study shows that experienced developers generally enjoy working with agents by \emph{controlling} agent behavior through strategic plans and supervision, rather than \emph{vibing}.
Below we discuss why experienced developers avoid vibing (\autoref{subsec:discussion-vibe}), actionable takeaways (\autoref{subsec:discussion-agent-use}), and future work (\autoref{subsec:discussion-future}).

\subsection{Why Don't Pros Vibe?}\label{subsec:discussion-vibe}

\topquote{I like working **with** agents. Not vibe coding. }{S96}

With agents capable of simpler tasks, developers can \emph{``focus on solving problems''} (S66) and architecture (S64), believing that agentic workflows will be too ubiquitous (\numStudyOne{4}, \numStudyTwo{13}) for them to \emph{``go back to writing code manually''} (S35) \emph{``EVER''} (S28).
Despite this optimism, experienced developers strategically \emph{control} agents rather than \emph{vibing}---refusing to let agents drive design or ``fully give in to the vibes...and forget that the code even exists''~\cite{karpathy2025vibecoding}.

We identify four reasons why experienced developers avoid coding with vibes:
(1) they value software engineering principles that are difficult to fully implant in agents (\numStudyOne{6});
(2) they work on production software with real stakeholders, not ``throwaway weekend projects''~\cite{karpathy2025vibecoding};
(3) pre-defined requirements in familiar codebases leave little room for exploratory coding, and delegating implementation to agents that lack the developer's rich context causes frustrating iterations;
and (4) when agentic solutions fail in unfamiliar domains, \emph{``it [could take] a lot of time to resolve them''} (S13).
These results imply that domain expertise---when available---supersedes vibes to drive the development of quality software.

\subsection{Generalizablity and Actionable Takeaways}
\label{subsec:discussion-agent-use}

\topquote{I never again want to give two [poops] about the specific best way to quijibo the toaster in dingledangle framework v0.21. The agent reads the latest docs and then is forced to comply.}{S88}



Although agentic tools have improved since our study---potentially shifting task suitability (\autoref{tab:suitability}), handling contexts better, and reducing the need for line-by-line review---we anticipate our core findings will persist. 
Specifically, production software quality matters and requires  human supervision and clear human-agent communication.
This aligns with \citet{Kam_2025}, who found that production-level agentic development relies on traditional software engineering skills, domain knowledge, and stakeholder alignment.

While we did not intend to develop a complete account of best practices, our findings recommend actionable strategies for developers' self-reflection.
First, prompt with explicit, detailed context (\autoref{fig:example_prompt}); while it is okay for prompts to be long, a vague prompt is likely to fail. 
For complex tasks, drafting explicit, multi-step plans in separate files is highly effective, provided the agent is constrained to implement the plan in manageable chunks (\eg, P1 and P2's 70+ step plans executed stepwise; \autoref{tab:prompt_size_and_verification} and \autoref{fig:example_prompt}). 
Second, do not expect to ``vibe.'' 
2025-era agents cannot \emph{autonomously} manage large software, requiring software engineering expertise to ensure code quality and project structure. 
Furthermore, developers can use \autoref{tab:suitability} to identify new use cases: while our participants avoided agents for complex tasks or core business logic, they considered agents potentially useful for scaffolding, test writing, refactoring, documentation, simple debugging, project setup, experiments, explanations, brainstorming, and knowledge lookup. 
Crucially, effective agent use requires practice (\emph{``the more experience I have with it, the more my personal workflow yields better results''} - S28). 
Thus, ``controversial'' tasks like planning might actually be suitable for developers who have honed their prompting skills. 
Finally, because initial AI failures hurt user trust more than successes build it~\cite{agentTrustPsychology}, developers adopting new workflows should expect early discouragement and keep iterating.

\subsection{Future Work}\label{subsec:discussion-future}

\topquote{This is the future of development, and it's very fun.}{S25}

We see two broad directions for future work: improvements to the interfaces for agentic coding, and further detailed studies of developer use of agents to derive best practices.

\mypara{Better Interfaces for Controlling Agents} 
One avenue for future work is improving agentic tools where they currently struggle, as our study revealed: automated testing in remote or realistic environments, tedious UI debugging through prompting, and understanding why an agent fails or hangs. 
Future research can design prompts and interfaces to improve these pain points. 
Similarly, because experienced developers plan in order to better control agent output, there is an opportunity to improve interfaces for the planning step. 
Careful interface design for planning might guide expert programmers inexperienced with AI into more successful workflows, and scaffold novices 
to prompt more effectively.

\mypara{Developing Best Practices}
To our knowledge, there is not yet any rigorously developed set of best practices for using coding agents,
although Anthropic~\cite{AnthropicBestPractices} and others~\cite{VibeEngineering,ClaudeCodeBeast} suggest some guidelines
and persistent \texttt{AGENT.md} instruction files can be a source of inspiration~\cite{FromCorrectnessToCollaboration2025,AgentReadmes2025}.
While our study produced an initial account of agent use by experienced developers, these patterns may not be optimal; as \citet{becker2025early2025} observed a 19\% coding performance drop with AI, it remains an open question whether those developers or our participants use agents to their full potential. 
To develop best practices, future work might focus on developers most successful with AI. 
Because productivity is notoriously hard to measure, researchers might recruit developers well-esteemed by their peers, or those with highly automated setups and a proven record of shipping working software. 
After gathering best practices from these ``most productive'' developers, further study is needed to validate that these practices \emph{actually} accelerate development in a greater variety of settings than \citet{becker2025early2025}.
Ultimately, we call for future work to confirm the most productive agentic usage patterns and verify that coding agents are measurably helpful for productivity.

\section{Related Work}
\label{sec:related}


Below we discuss AI agents for software development, recount usability studies of LLMs for coding, and compare recent empirical investigations of vibe coding in particular.

\subsection{Agents for Coding}

With transformer advancements~\cite{AttentionIsAllYouNeed} and abundant open-source training data, LLMs quickly entered programming. 
They debuted as super-charged autocomplete via GitHub Copilot~\cite{GitHubCopilot} powered by OpenAI's Codex~\cite{openai2021codex} in 2021, evolving into conversational partners with instruction-following models~\cite{InstructGPT} like ChatGPT in 2022~\cite{ChatGPT}.

LLMs improved steadily with the ability to think aloud~\cite{ChainOfThoughtPrompting} and autonomously plan and enact multi-step workflows~\cite{AutoGPT}. 
Their coding ability reached a milestone with Devin in 2024~\cite{wu2024introducingDevin}, ``the first AI software engineer,'' soon followed by academia's SWE-agent~\cite{SWEagent} targeting ``end-to-end software engineering.''
Moving beyond local autocomplete, these systems were \emph{agentic}: autonomously taking step-by-step actions to read, modify, and test whole codebases by iteratively invoking tools (\eg search, read, edit, run) and fixing errors like a human's edit-run-debug cycle.


Beyond broadly capable agents like Devin and SWE-agent, specialized AI agents appear for nearly every software development phase, including requirements engineering, debugging, and testing~\citep{liu2024agentsSWE,jin2024agentsSWE}.
However, widely deployed systems (Cursor, GitHub Copilot Agent, Claude Code, Codex CLI) default to broadly capable models (rather than task specific ones), though users can still prompt specialized subagents (\eg Claude Code Subagents~\cite{ClaudeCodeSubagents}). 
While this autonomous paradigm first drove significant progress~\cite{SWEagent} on SWE-bench~\cite{SWEbench}, autonomous choices may not be necessary for benchmarks: a carefully crafted non-agentic pipeline of fixed steps performs competitively with state-of-the-art agents~\cite{xia2024agentless}. 
This counter-result---effectively extreme prompt engineering---highlights that prompting strategy heavily dictates success with LLMs. We thus expect large variations in user success based on prompting expertise. 
Below we discuss usability research on LLMs for coding.

\subsection{Usability of LLMs for Coding}


\mypara{Prompt Quality}
In the studies of LLM use for coding,
the correlation between prompt quality and success is a repeated theme in both the autocomplete and the agentic eras.
In the autocomplete era, prompts may be robust to word choice but success drops dramatically if the prompt is missing key information~\cite{lucchetti2025substanceBeatsStyle}, surveyed developers thus report ``clear explanations'' as the most common prompting strategy~\cite{liang2024survey}, and surveyed students similarly suggest that output quality is improved by using specific prompts, decomposing problems into steps, and iteratively improving a prompt~\cite{akhoroz2025students}.
Expertise matters: non-programmers cannot prompt as successfully as even novice programmers in a lab environment~\cite{NonExpertGenAI}.
%
Similar patterns hold in the agentic era.
In lab tasks, \citet{eibl2025challengesOpportunities} found key information was left out of 65\% of prompts,
and in a retrospective of an agentic system for resolving real-world software issues, \citet{takerngsaksiri2025hula} note their number one takeaway is that LLM helpfulness ``heavily relies on a detailed input description.''
Participants in our study as well highlighted that prompts should include clear context and explicit instructions (\numStudyOne{12}, \numStudyTwo{43}).
Prompt quality is a key bottleneck in LLM usability.


\mypara{AI Utility}
At an even broader view, there is the question of whether or when LLMs are measurably helpful to programmers, as the extra burdens of prompting and code review may overwhelm any gains.

From the autocomplete era, quantitative studies of this question differ in their results.
A couple of controlled lab studies showed positive results, those with Copilot were 56\% faster on a single task in \citet{peng2023copilotProductivity} (n=95, mid-2022), and when experienced Python developers were asked to implement API endpoints for a web server, those with Copilot were much more likely to submit solutions that passed unit tests (60.5\% versus 39.5\%, p < 0.01, n=202)~\cite{bauer2024copilotQuality}.
But on the three lab tasks in \citet{ExpectationVsExperienceCodeGenAI}, participants with Copilot were broadly similar in success rate and speed compared to those without (n=24, $\le$ Jan 2022), although participants preferred to use Copilot; the authors suspect the lack of measurable gain is because of buggy generated code leading to an increase in debugging time, with participants generally underestimating the effort to fix generated code.
In a private retrospective analysis of bug rate and pull request (PR) throughput among a large sample of real-world developers, those with access to Copilot produced more bugs than those without, without any notable difference in PR throughput~\cite{uplevel2024}.

In the agentic era, quantitative support for gains is similarly mixed. Developers believe agentic AI is helpful:
of those using agents, 69\% believe it has increased their productivity~\cite{stackoverflow2025survey}.
But \citet{becker2025early2025} found otherwise: they studied 16 experienced open source maintainers, allowing them to use or not use AI tools on their projects (condition randomized per-task, n=246 tasks). While these developers estimated that AI access made them 20\% faster, on the whole they were in fact 19\% slower when they could use AI. The non-helpfulness may be because this setup was a worse-case scenario for AI: large, complex repositories, where developers are intimately familiar with the code and have tacit knowledge unavailable to the AI, while the developers are also over-optimistic about AI utility~\cite{becker2025early2025}. The study measured a low accept rate for code generations of <50\%, which accords with the 25\% accept rate of the agentic tool in \citet{takerngsaksiri2025hula}.
In roughly the same time period (first half of 2025), analyses of agent-generated PRs on GitHub show higher accept rates, 38\%-65\%~\cite{AIDevCorpus} or 84\%~\cite{Watanabe2026GitHubPRs}, although still lower than for human-generated PRs: 77\%~\cite{AIDevCorpus} and 91\%~\cite{Watanabe2026GitHubPRs} respectively.
However, in an industry setting, some evidence of utility was found by \citet{sarkar2026ai}: 24 software companies that adopted agents saw a 39\% increase in pull request merge rate once agents became the default in Cursor, compared to 8 companies not using Cursor, without any measurable change in PR sizes, merge reversals, or bugfix mergers.
WhatsApp also reported hodgepodge but measurable benefits from incorporating agentic AI in their development workflow~\cite{WhatsCode}.
%
With these mixed results, work is needed to understand when and why AI tools actually help.
We list scenarios where experienced developers prefer using agents (\autoref{tab:suitability}) but  controlled measurement of scenario-specific workflow speedup remains future work.

\mypara{Human-Agent Collaboration}
To better understand the mixed results on AI in software development, recent qualitative studies probed into the collaborative dynamics between humans and agentic AI~\cite{kumar2025why, dhanorkar2026human,feng2026chi}.
Though these studies focused on the general developer population, they all revealed the experienced developers' preference for (and challenges with) control alongside strategies like decomposition and providing context, and the perceptions of ``agents as source of collaboration''.
The closest to our work, \citet{kumar2025why} further hinted at some tasks that did not suit the agents (for ones that require domain knowledge).
Our work supports the above core finding that experienced developers control agents as opposed to giving in, and enriches our understanding in two ways.
First, we provided a fine-grained analysis of various prompting strategies and verification tactics among experienced developers (Tables \ref{tab:prompt_size_and_verification} and \ref{tab:prompt_context}) that led to success in controlling the agents.
Second, we identified a large set of tasks that are suitable for the agents (or otherwise) that could further explain the rationale for control instead of full delegation.

\subsection{Empirical Studies of Vibe Coding}

As LLMs advance to generate entire codebases, developers might disengage from raw code to focus on prompting. \citet{karpathy2025vibecoding} dubbed this ``vibe coding,'' where developers ``fully give in to the vibes, embrace exponentials, and forget that the code even exists.'' 
This raises the tantalizing possibility that formal code could become incidental to software creation, potentially enabling even non-programmers to develop software.
%
%
Definitions of vibe coding vary~\cite{BuildingSoftwareByRollingTheDice,pimenova2025goodVibrations}, but its core feature seems to be trusting the AI to achieve ``flow and joy''~\cite{pimenova2025goodVibrations}, which (in our interpretation) often leads vibe coders to avoid manual review that kills the vibes. 
While practitioners are generally aware of the resulting lower code quality and restrict vibing to specific use cases~\cite{pimenova2025goodVibrations,fawzy2025vibeCoding} \eg ``throwaway weekend projects''~\cite{karpathy2025vibecoding}, \citet{fawzy2025vibeCoding} warn that AI hype may turn non-programmers into ``vulnerable developers'' managing real-world software without the requisite expertise. 
Analyzing livestreamed vibe coding sessions, \citet{BuildingSoftwareByRollingTheDice} observed that while participants enjoyed agentic workflows, they could suffer from design fixation, spent significant time waiting for AI responses (over 20\% of livestream time), and engaged in redundant ``dice rolling'' by repeatedly reissuing the same prompts.

Crucially, empirical evidence suggests vibe coding still demands programming expertise. 
\citet{BuildingSoftwareByRollingTheDice} noted non-professionals struggled to debug, with one user blindly copy-pasting 31 consecutive error messages as debugging prompts. 
Conversely, \citet{sarkar2025vibe} found that experienced developers rely on their expertise to strategically review code, to switch to manual edits, and to recover ``when the vibes are off.''
\citet{chandrasekaran2025productionGrade} also found that creating production-grade apps requires a non-vibey, disciplined approach with ``deliberate oversight.''
An expertise gap is also evident in students: advanced software engineering students provide more context and interact more with the code than novices~\cite{geng2025studentVibe}. 
Consequently, as of 2025, non-professionals will likely struggle to produce complex software with pure vibes.

Thus, the experienced developers we observed avoided pure vibe coding. 
When implementing new features, they actively maintained control by planning (manually or by revising AI drafts) and meticulously reviewing output.
Specifically, 69\% (9/13) of observed developers carefully reviewed every change, aligning with the open-source developers \citet{becker2025early2025} studied, where 75\% said they read \emph{every} line of AI-generated code.
Echoing \citet{BuildingSoftwareByRollingTheDice}, our participants also prompted with by referencing various kinds of contexts, used quality assurance strategies like testing, and enjoyed collaborating with agents.
While advancing LLMs may eventually make vibe coding the future of software development---even for non-programmers---for complex, production-grade software, it is not the present.

\section{Conclusion}
\label{sec:conclusion}


\topquote{I think AI agents are amazing as long as you are the driver \& reviewing its work.}{S64}

To gain insight into the current practice of agentic coding by experts,
we studied developers with 3+ years of professional experience through
13 field observations and a qualitative survey of 99 respondents. We aimed to understand their values, workflow strategies, task suitability, and sentiments when using coding agents.
We find that experienced developers do not currently vibe code.
Instead, they carefully \emph{control} the agents through planning and active supervision because they care about software quality.
Although they may not always read agent-generated code, they are careful not to lose control and do not let agents run completely autonomously, particularly having the agents only work on a few tasks at a time.
They generally enjoy using agents as a source of collaboration---not delegation---finding them suitable for accelerating straightforward, repetitive, and scaffolding tasks;
and yet opinions are mixed about writing plans with agents, and they do not use agents for core business logic or complex tasks.
AI capabilities and interfaces are changing fast: this work paints a picture of what is and is not working now, serving as a reference point both to calibrate expectations about current AI and to anchor comparison in future years.
As of now, agents are not taking over yet---experienced developers are still in control.



\begin{acks}
Our thanks to Bryan Min, Saketh Kasibatla, and Matthew Beaudouin-Lafon for helpful feedback on the study design and recruitment.
This work was supported by U.S. National Science Foundation Grants No. 2432644 (\emph{Direct Manipulation for Everyday Programming}) and No. 2107397 (\emph{Human-Centric Program Synthesis}).
\end{acks}
\notarxiv{\section{Data-Availability Statement}

An artifact for this paper includes redacted versions of our study materials, raw analysis results, and a supplementary appendix,  available at \packageURL (a public, anonymous OSF project; OSF produces DOIs on de-anonymization for camera-ready).
We are unable to share raw data given our institutional review board protocol and participant consent.

 }

\balance

\bibliographystyle{ACM-Reference-Format}
\bibliography{references}

\clearpage
\onecolumn

\appendix

\setcounter{figure}{0}
\setcounter{table}{0}
\renewcommand{\thefigure}{A\arabic{figure}}
\renewcommand{\thetable}{A\arabic{table}}

\section{Appendix}
\label{sec:appendix}

Below are a full list of the GitHub repositories used to invite survey respondents (\autoref{sec:survey-recruitment}), non-demographic survey questions (\autoref{sec:survey-questions}), and quotes from survey respondents about prompt strategies (\autoref{sec:survey_prompting_quotes}), and task suitability (\autoref{sec:task_suitability_full}).

\subsection{Survey Recruitment Sources}
\label{sec:survey-recruitment}

\begin{table}[h]
\centering
\footnotesize
\setlength{\tabcolsep}{3pt}
\caption{GitHub repositories from which we invited participants by querying recent commits, pull requests, issues, stargazers, and forkers within the past year.}
\label{tab:repositories}
\begin{tabularx}{\linewidth}{lX}
\toprule
\textbf{Category} & \textbf{Repository} \\
\midrule
\textit{AI/ML Frameworks \& Tools} & All-Hands-AI/OpenHands, HKUDS/DeepCode, OpenBMB/ChatDev, Significant-Gravitas/AutoGPT, aider-ai/aider, anthropics/claude-code, gpt-engineer-org/gpt-engineer, microsoft/autogen, microsoft/guidance, TabbyML/tabby, TransformerOptimus/SuperAGI, crewAIInc/crewAI, joaomdmoura/crewAI \\
\midrule
\textit{Development Tools \& IDEs} & cursor/cursor, cline/cline, stackblitz/bolt.new, Kilo-Org/kilocode \\
\midrule
\textit{Language Models \& AI Research} & EleutherAI/gpt-neo, EleutherAI/gpt-neox, facebookresearch/llama, google-research/bert, openai/CLIP, openai/gpt-2, openai/whisper, microsoft/CodeBERT, salesforce/CodeT5 \\
\midrule
\textit{Data Science \& Libraries} & numpy/numpy, pandas-dev/pandas, scikit-learn/scikit-learn, plotly/plotly.py, chroma-core/chroma \\
\midrule
\textit{Development Frameworks} & continuedev/continue, jerryjliu/llama\_index, run-llama/llama\_index, vercel/ai, torantulino/auto-gpt, yoheinakajima/babyagi \\
\midrule
\textit{Microsoft Development Tools} & microsoft/JARVIS, microsoft/pylance, microsoft/unilm \\
\midrule
\textit{Other Development Tools} & jupyter/notebook, google-gemini/gemini-cli, openai/chatgpt-retrieval-plugin \\
\bottomrule
\end{tabularx}
\end{table}

\subsection{Survey Questions}
\label{sec:survey-questions}

\begin{figure}[h]
\relsize{-2}
\begin{tcolorbox}[colback=white,colframe=black!25,arc=1.5pt,boxrule=0.3pt,left=0pt,right=0pt,top=0pt,bottom=0pt]
\rowcolors{1}{}{gray!10}
\setlength{\tabcolsep}{3pt} 
\renewcommand{\arraystretch}{1.45} 
\newcommand{\nextrowwhite}[0]{\global\rownum=0\relax}
\nextrowwhite
\begin{tabular}{p{0.985\textwidth}}
\showrowcolors
\textbf{To guide you in answering the following questions, think about the last time you used AI agents to help develop software. Then, answer the following questions based on that experience.}\nextrowwhite \\
1. What was your task? \\
2. What are three important things you care about when developing software? \\
3. Which of these did you care about \textbf{most} when using AI agents to complete this task? If none of the above, what did you care about? \\
4. What tool(s) did you use to work with the agents? Select all that apply. \textit{[7 options + ``Other'' free response, see \autoref{fig:survey_tasks_experience_tools_desired_attributes}c]} \\
5. Why did you choose the above tool(s) for this task? \\
6. How did you prompt—was there special information you included or any prompting tricks or prompt engineering tactics you used? \\
7a. For this task, did you use one agent at a time, or did you run multiple in parallel? \textit{[2 options]} \\
7b. If you used parallel agents, how did you set that up? \\
8. For this task, how frequently did you feel like you had to modify or change the code output generated by the agent(s)? \textit{[5 options]} \\
9. What part(s) of the task did the agent(s) perform well on? Poorly on? \\
10. After finishing (or abandoning!) the task, how suitable do you feel agent(s) are for the task? \textit{[6 options]} \\
11a. For this task, how much did you enjoy developing software with agents compared to without? \textit{[6 options]} \\
11b. Can you explain your rating for the previous question?\nextrowwhite \\
\arrayrulecolor{black!25}\midrule
\textbf{Answer the following questions thinking about your \textit{overall experiences this year} in software development.}\nextrowwhite \\
12. Thinking of all the responsibilities you have as a software developer, which kinds of tasks do you prefer to perform \textbf{without assistance} from agents? \\
13. Thinking of all the responsibilities you have as a software developer, which kinds of tasks do you prefer to perform \textbf{with} agents? \\
Optional: Do you have any other comments or anything else you would like us to know? \\
\end{tabular}
\end{tcolorbox}
  \caption{Non-demographic questions from survey. Unless the number of response options are noted, questions are free response.}
  \label{fig:survey_questions}
\end{figure}

\newpage

\subsection{Survey Quotes About Prompting and Task Suitability}
\label{sec:survey_prompting_quotes}

\begin{figure}[h]
\relsize{-2}
\begin{tcolorbox}[colback=white,colframe=black!25,arc=1.5pt,boxrule=0.3pt,left=0pt,right=0pt,top=0pt,bottom=0pt]
\rowcolors{1}{}{gray!10}
\setlength{\tabcolsep}{3pt} 
\renewcommand{\arraystretch}{1.45} 
\begin{tabular}{p{0.985\textwidth}}
\showrowcolors
Usually, I just provide them with the \textbf{task details}, but sometimes I need to specify the steps. (S21) \\
I give \textbf{clear instructions} and ask for specific output format (S23) \\
I first had it \textbf{help me develop a "whitepaper"} describing the project and planning out the development in phases (that could be independently evaluated).  Then I had it execute the phases. (S38) \\
It needed some iteration to understand the code and \textbf{I provided more guidelines and it got better and better}. (S43) \\
IF given \textbf{clear instructions} they do my work (S44) \\
I do use a lot of prompt tricks, but overally just good comunication and explaining in detail. Like, \textbf{overexplaining in a LOT of detail}. (S49) \\
Very tailored prompting. Needs a coding protocol. Needs an SDLC [software development life cycle] protocol. \textbf{Lots of context} on code structure, includes PRD, includes coding plan, includes architecture plan, HITL [human in the loop] after each task to avoid drift. (S52) \\
Usually I give brief explanation of the task on the top, followed by \textbf{a list of details}. Also I try to write prompts in Markdown format. (S53) \\
Please create new feature doing x and y based on base class BackgroundProcess from bin/BackgroundProcess.php and \textbf{example implementation} featureA from bin/featureA.php (S57) \\
The prompts are always very specific, with \textbf{as much info as possible} for example - Filenames, function names, variable names, error messages etc. I focus on ensuring prompt has clear statements on what I [want] the LLM to look at and what I want it to do. Small and Specific modifications (S59) \\
(1) Defined \textbf{clear instructions} first. Split to multiple files for maintainability, \newline
(2) Brainstorming first, then start the task, \newline
(3) Refer to files directly [in] the prompt to provide more context if needed (S60) \\
Before starting a task I give a very thorough prompt of the requirements and context \& \textbf{I ask for explicit clarification} if any details aren't clear to ensure the agent is on the same page as me. (S64) \\
As someone who values scalability and security, I make sure to cover \textbf{every single edge case} I can think of. I also provide the agent with any sample resources I need it to use for reference, such as code structures and style guides...the agent performs well when it receives \textbf{detailed prompts}. (S66) \\
My prompts were simply about stating the requirements clearly and explaining the problem \textbf{thoroughly}. I found that \textbf{longer prompts} tend to work better than shorter ones (S72) \\
Always \textbf{treat AI as a smart kid that knows nothing about you and outside world} (more like introvert). So you must write the context, write strict rules (do's and don't), write the guides, and finally write the task. (S75) \\
1. Direct instruction following \newline
2. Test driven approach \newline
3. Dividing big projects into small tasks and creating \textbf{todo lists} (S76) \\
I maintain a CLAUDE.local.md file with project context and development principles, use descriptive commit messages for context, and leverage git worktrees for parallel development. I focus on \textbf{clear, specific requests with relevant file paths and context}, often referencing recent commits and PRs to provide the AI with understanding of the current work (S85) \\
If you have \textbf{clear plan}...would save you much time (S89) \\
If I provide the right context or my \textbf{exact requirement} or function signature, I get more than a 80\% success rate. (S92) \\
\end{tabular}
\end{tcolorbox}
\caption{Select quotes from survey respondents who found agents performing well at \taskSU{following well-defined plans}{28}{2}. A common theme is providing clear instructions with as much information as possible. Bolded emphasis is ours.}
\label{fig:survey_prompting_quotes}
\end{figure}

\newpage

\subsection{Task Suitability, With Representative Quotes}
\label{sec:task_suitability_full}

\begin{table}[b]
\vspace{-2em}
\caption{Agentic task suitability from the surveys. The 59 tasks below were mentioned in at least 5 of the 99 surveys. Tasks are grouped into ``Maybe suitable'', ``Controversial'', and ``Maybe unsuitable'' based on the ratio of responses (imbalance of $\ge$ 2.5:1 to be ``suitable'' and $\le$ 1:2.5 to be ``unsuitable''). Note that ``Refactoring'' and ``Debugging'' are subdivided into simple/general/complex, although complex debugging did not receive five mentions and is not reported.
Bars depict the number of survey responses indicating \textcolor{scolor}{suitablility (S)} or \textcolor{ucolor}{unsuitablility (U)}, \SU{S}{U} are the counts, and the final column gives example survey quotes.
}
\label{tab:suitability_with_evidence}
\vspace{-.5em} 
{
\rowcolors{1}{}{gray!10}
\fontsize{7.25pt}{7.75pt}\selectfont 
\centering
\setlength{\tabcolsep}{1.7pt} 
\newcommand{\barscale}{0.8}
\newcommand{\SUbar}[2]{\pgfmathparse{\barscale*23-\barscale*#2}\hspace{\pgfmathresult pt}\raisebox{-0.31em}{\tikz \fill[ucolorlight] (0,0) rectangle (\barscale*#2pt, 1.1em);\tikz \fill[scolor] (0,0) rectangle (\barscale*#1pt, 1.1em);}}
\centerline{\begin{tabular}{r l c l} 
    \hiderowcolors
    \toprule
    
    \textbf{\relsize{1}Task or Use Case} &
\hspace{4pt}\textbf{\relsize{1}\textcolor{ucolor}{\clipbox{0 0 3pt 0}{$\leftarrow$}\,U}} \textbf{\relsize{1}\textcolor{scolor}{S\,\clipbox{3pt 0 0 0}{$\rightarrow$}}} &
    \textbf{\relsize{1}\SU{S}{U}}
    & \textbf{\relsize{1}Evidence Example(s)} \\
    \hiderowcolors
    \midrule
    \multicolumn{4}{l}{\global\rownum=0\relax\textbf{\relsize{1}Maybe suitable for...}} \\ 
    \midrule
    \showrowcolors
Accelerating productivity & \SUbar{35}{2} & \SU{35}{2} &\hspace{-1pt}\emph{``Saves a significant amount of time and improves my efficiency''} (S21)\\
Small/simple/straightforward tasks & \SUbar{33}{1} & \SU{33}{1} &Prefer for \hspace{-1pt}\emph{``the grunt work''} (S25), \hspace{-1pt}\emph{``tasks that don't require...innovations''} (S30)\\
Following well-defined plans & \SUbar{28}{2} & \SU{28}{2} &Prefer for \hspace{-1pt}\emph{``concrete, well-defined, and implementation-focused''} tasks (S73)\\
Generating new code & \SUbar{27}{2} & \SU{27}{2} &Prefer for \hspace{-1pt}\emph{``coding''} (S41); \hspace{-1pt}\emph{``quickly write code based on [my] prompts''} (S50)\\
Tedious/repetitive tasks & \SUbar{26}{0} & \SU{26}{0} &Prefer for \hspace{-1pt}\emph{``reptitive [tasks]''} (S65), \hspace{-1pt}\emph{``[tasks] I can do without even thinking''} (S66)\\
Scaffolding or boilerplate & \SUbar{25}{0} & \SU{25}{0} &Prefer for \hspace{-1pt}\emph{``writing outline of the code''} (S53); agent did well at \hspace{-1pt}\emph{``boilerplate''} (S60)\\
Writing tests & \SUbar{19}{2} & \SU{19}{2} &Agent did \hspace{-1pt}\emph{``well on writing simple tests''} (S74)\\
Refactoring (general) or code improvement & \SUbar{18}{3} & \SU{18}{3} &\hspace{-1pt}\emph{``Refactoring''} (S16,S18,S61,S68,S76,S85); \hspace{-1pt}\emph{``restructuring code''} (S65)\\
All or most all tasks & \SUbar{20}{0} & \SU{20}{0} &Which...do you prefer to perform with agents? \hspace{-1pt}\emph{``Everything''} (S49)\\
Writing/updating documentation & \SUbar{20}{0} & \SU{20}{0} &Prefer for \hspace{-1pt}\emph{``writing documentation''} (S64)\\
Backend development & \SUbar{14}{1} & \SU{14}{1} &\hspace{-1pt}\emph{``Software Engineering Backend''}, agent \hspace{-1pt}\emph{``Suitable''}, user \hspace{-1pt}\emph{``Moderately pleased''} (S98)\\
Debugging (simple) or simple fixes & \SUbar{12}{3} & \SU{12}{3} &Prefer for \hspace{-1pt}\emph{``bug fixes which I already `know' what the solution is''} (S64)\\
Refactoring (simple) or modifying code or cleanup & \SUbar{14}{0} & \SU{14}{0} &Prefer for \hspace{-1pt}\emph{``code changes''} (S9) or \hspace{-1pt}\emph{``long Typescript cleanup sessions''} (S96)\\
Starting or setting up a new project & \SUbar{10}{4} & \SU{10}{4} &Prefer for \hspace{-1pt}\emph{``green field development''} (S47) or \hspace{-1pt}\emph{``creating initial folder structure''} (S61)\\
Frontend development & \SUbar{10}{3} & \SU{10}{3} &Did well at \hspace{-1pt}\emph{``generating flutter...code following instructions''} (S60)\\
Prototyping or small projects & \SUbar{12}{0} & \SU{12}{0} &Did well on \hspace{-1pt}\emph{``first prototypes''} (S27) and \hspace{-1pt}\emph{``simple small projects''} (S77)\\
Explaining/analyzing code/APIs/errors & \SUbar{10}{1} & \SU{10}{1} &\hspace{-1pt}\emph{``[creating] a code map''} (S22); good at \hspace{-1pt}\emph{``explaining how something works''} (S59)\\
Exploring alternatives or experimenting & \SUbar{7}{0} & \SU{7}{0} &Prefer for \hspace{-1pt}\emph{``throwaway demos/experiments''} (S24), \hspace{-1pt}\emph{``architecture exploration''} (S47)\\
Helping the developer learn & \SUbar{6}{1} & \SU{6}{1} &\hspace{-1pt}\emph{``Code explanations are helping me learn new things''} (S59)\\
Looking up knowledge for the developer & \SUbar{6}{1} & \SU{6}{1} &\hspace{-1pt}\emph{``Helpful for research''} (S25); prefer for \hspace{-1pt}\emph{``learning \& knowledge lookup''} (S76)\\
Data processing/analysis & \SUbar{5}{2} & \SU{5}{2} &Did well at \hspace{-1pt}\emph{``creating helpers to process data''} (S19)\\
Collaboratively talking out problems & \SUbar{6}{0} & \SU{6}{0} &\hspace{-1pt}\emph{``I really like bouncing ideas off AI to see what things it thinks of that I didn't.''} (S7)\\
Writing helper functions & \SUbar{6}{0} & \SU{6}{0} &\hspace{-1pt}\emph{``It is good at generating...utility functions''} (S55)\\
Testing & \SUbar{5}{0} & \SU{5}{0} &\hspace{-1pt}\emph{``Good at loop iteration on running - fixing tests''} (S18)\\
LLM following a knowledge base & \SUbar{4}{1} & \SU{4}{1} &Did well at \hspace{-1pt}\emph{``understanding context from extensive knowledge base''} (S85)\\
Using major/common frameworks/libraries & \SUbar{4}{1} & \SU{4}{1} &\hspace{-1pt}\emph{``Batteries all included frameworks...have a bizarrely powerful synergy with AI''} (S88)\\
    \hiderowcolors
    \midrule
    \multicolumn{4}{l}{\global\rownum=0\relax\textbf{\relsize{1}Controversial}} \\ 
    \midrule
    \showrowcolors
High level plans (\eg project or architecture design) & \SUbar{13}{23} & \SU{13}{23} &Dislike for \hspace{-1pt}\emph{```big picture' design''} (S9); S28 uses AI for \hspace{-1pt}\emph{``bigger design and planning''}\\
Debugging (general) & \SUbar{12}{8} & \SU{12}{8} &Did well on \hspace{-1pt}\emph{``analy[zing] bugs''} (S36); \hspace{-1pt}\emph{``no LLMs [debugged] well''} (S57)\\
Brainstorming or conceptualization & \SUbar{11}{5} & \SU{11}{5} &\hspace{-1pt}\emph{``For a new problem, I just engage with some LLM, get ideas''} (S92)\\
UI/HTML/CSS & \SUbar{9}{4} & \SU{9}{4} &\hspace{-1pt}\emph{``It [saves] time to do UI layout''} (S26); dislike for \hspace{-1pt}\emph{``fine tunning CSS/HTML''} (S23)\\
Understanding project architecture & \SUbar{6}{7} & \SU{6}{7} &\hspace{-1pt}\emph{``Confusion on environment''} (S52); use \hspace{-1pt}\emph{``to understand...sections in...code base''} (S59)\\
Long context sessions or between-task consistency & \SUbar{4}{8} & \SU{4}{8} &\hspace{-1pt}\emph{``Memory...doesn't do well. Context management is not great sometimes.''} (S17)\\
Code review & \SUbar{7}{4} & \SU{7}{4} &Prefer for \hspace{-1pt}\emph{``PR reviews''} (S9); dislike for \hspace{-1pt}\emph{``reviews and quality assurance''} (S31)\\
Writing feature-level implementation plans & \SUbar{6}{3} & \SU{6}{3} &Prefer for \hspace{-1pt}\emph{``PRD generation''} (S55); \hspace{-1pt}\emph{``planning...went off the rails frequently''} (\newS{100})\\
Creative vision or creative tasks & \SUbar{4}{4} & \SU{4}{4} &Did well on \hspace{-1pt}\emph{``content generation''} (S41); dislike for \hspace{-1pt}\emph{``tasks [needing] creativity''} (S56)\\
Understanding task context & \SUbar{3}{4} & \SU{3}{4} &Poor at \hspace{-1pt}\emph{``reading GitHub issues''} (S74); \hspace{-1pt}\emph{``excels when provided with...context''} (S85)\\
Domains/libraries unfamiliar to the developer & \SUbar{4}{2} & \SU{4}{2} &\hspace{-1pt}\emph{``For an unknown task...it gives a great headstart to look into solutions.''} (S51)\\
Creating configurations & \SUbar{3}{3} & \SU{3}{3} &Dislike for \hspace{-1pt}\emph{``Ansible dev''} (S29); prefer for \hspace{-1pt}\emph{``creating configurations''} (S34)\\
Version control & \SUbar{2}{4} & \SU{2}{4} &Dislike for \hspace{-1pt}\emph{``code setup in git''} (S17); \hspace{-1pt}\emph{``git commands I mostly do myself''} (S79)\\
Handling many files at once & \SUbar{2}{3} & \SU{2}{3} &\hspace{-1pt}\emph{``It poorly perfom when the codebase is modularized''} (S75)\\
    \hiderowcolors
    \midrule
    \multicolumn{4}{l}{\global\rownum=0\relax\textbf{\relsize{1}Maybe unsuitable for...}} \\ 
    \midrule
    \showrowcolors
One-shotting code without modification/verification & \SUbar{5}{23} & \SU{5}{23} &Did \hspace{-1pt}\emph{``poorly on...clean, production-ready code without my manual refactoring''} (S31)\\
Integrating with existing code; legacy code & \SUbar{3}{17} & \SU{3}{17} &Did \hspace{-1pt}\emph{``poorly on adding new features to existing functionality''} (S33)\\
Complex tasks (not necessarily big) & \SUbar{3}{16} & \SU{3}{16} &Dislike for \hspace{-1pt}\emph{``heavily logical''} tasks (S78); or \hspace{-1pt}\emph{``implementing new complex logic''} (S61)\\
Business logic or tasks requiring domain knowledge & \SUbar{2}{15} & \SU{2}{15} &Dislike \hspace{-1pt}\emph{``when I start a customer software tailored with business logic''} (S45)\\
Writing performant code & \SUbar{3}{9} & \SU{3}{9} &Poor on \hspace{-1pt}\emph{``performance optimization for high-load scenarios''} (S91)\\
Replacing human expertise or decision making & \SUbar{0}{12} & \SU{0}{12} &\hspace{-1pt}\emph{``I like coding alongside agents. Not vibe coding. But working **with**''} (S96)\\
Handling CI/deployment infrastructure & \SUbar{3}{8} & \SU{3}{8} &Dislike for \hspace{-1pt}\emph{``deployment''} (\newS{103}); or for \hspace{-1pt}\emph{``CI/CD pipeline configuration''} (S85)\\
Handling details & \SUbar{2}{7} & \SU{2}{7} &\hspace{-1pt}\emph{``Poor performance in handling my data scenario requirements''} (S11)\\
Big tasks (not necessarily complex) & \SUbar{1}{7} & \SU{1}{7} &\hspace{-1pt}\emph{``It struggled with large-scale code generation and refactoring.''} (S73)\\
High-stakes or privacy-sensitive tasks & \SUbar{0}{8} & \SU{0}{8} &Dislike for \hspace{-1pt}\emph{``high-risk codes''} (S53); or for \hspace{-1pt}\emph{``handling sensitive...codebases''} (S58)\\
Security-critical code & \SUbar{2}{5} & \SU{2}{5} &Dislike for \hspace{-1pt}\emph{``APIs keys''} (S12) or \hspace{-1pt}\emph{``security-critical code reviews''} (S85)\\
Vague/unclear/open-ended tasks & \SUbar{0}{7} & \SU{0}{7} &Poor with \hspace{-1pt}\emph{``abstracted prompt''} (S30); or \hspace{-1pt}\emph{``sweeping changes...without a plan''} (S59)\\
Using uncommon or private libraries/languages & \SUbar{1}{5} & \SU{1}{5} &\hspace{-1pt}\emph{``Some internal platforms I use at work that agents don't know much about''} (S8)\\
Complex logic & \SUbar{0}{6} & \SU{0}{6} &Poor at \hspace{-1pt}\emph{``complex multi-step business logic without [my guidance]''} (S31)\\
Choosing technologies/frameworks/libraries & \SUbar{1}{4} & \SU{1}{4} &Dislike for \hspace{-1pt}\emph{``hosting and technology/framework/library decisions''} (S42)\\
Refactoring (complex) or improving architecture & \SUbar{1}{4} & \SU{1}{4} &Dislike for \hspace{-1pt}\emph{``core changes to the structure of the project''} (S16)\\
Tasks highly familiar to the developer & \SUbar{1}{4} & \SU{1}{4} &Dislike for \hspace{-1pt}\emph{``few line changes with deep domain knowledge''} (S94)\\
Database design/management & \SUbar{0}{5} & \SU{0}{5} &Dislike for \hspace{-1pt}\emph{``database design''} (S42), \hspace{-1pt}\emph{``database management''} (S50)\\
New API versions after LLM's knowledge cutoff & \SUbar{0}{5} & \SU{0}{5} &\hspace{-1pt}\emph{``Specific Azure APIs were not used [probably] due to API changes''} (S32)\\
    \bottomrule
\end{tabular}}

}

\end{table}

\end{document}